\documentclass[%
aip,
amsmath,amssymb,
reprint,%
]{revtex4-1}

\usepackage{graphicx,float}
\usepackage{dcolumn}
\usepackage{bm}
\usepackage{physics}
\usepackage{pgfplots}
\usepackage[utf8]{inputenc}
\usepackage[T1]{fontenc}
\usepackage{mathptmx}
\usepackage{color,soul}
\usepackage{cancel}

\begin{document}
	
	\preprint{AIP/123-QED}
	
	\title[]{Benchmarking an Improved Statistical Adiabatic Channel Model for Competing Inelastic and Reactive Processes}
	
	\author{Maarten Konings}
	\email{maarten.konings@kuleuven.be.}
	\affiliation{ 
		KU Leuven, Division of Quantum Chemistry and Physical Chemistry, Department of Chemistry, Celestijnenlaan 200F, 3001 Leuven, Belgium.
	}
	\author{Benjamin Desrousseaux}%
	\affiliation{
		Universit\'{e} de Rennes 1, CNRS, IPR (Institut de Physique de Rennes) - UMR 6251, F-35000 Rennes, France.
	}
	\author{Fran\c{c}ois Lique}%
	\affiliation{
		Universit\'{e} de Rennes 1, CNRS, IPR (Institut de Physique de Rennes) - UMR 6251, F-35000 Rennes, France.
	}
	\author{J\'{e}r\^{o}me Loreau}%
	\affiliation{ 
		KU Leuven, Division of Quantum Chemistry and Physical Chemistry, Department of Chemistry, Celestijnenlaan 200F, 3001 Leuven, Belgium.
	}%
	\date{23 August 2021}
	
	\begin{abstract}
		Inelastic collisions and elementary chemical reactions proceeding through the formation and subsequent decay of an intermediate collision complex, with associated deep well on the potential energy surface, 
		pose a challenge for accurate fully quantum mechanical approaches, such as the close-coupling method.
		In this study, we report on the theoretical prediction of temperature-dependent state-to-state 
		rate coefficients for these complex-mode 
		processes, using a 
		statistical quantum method. This statistical adiabatic channel model is benchmarked by direct comparison using accurate rate coefficients from the literature for a number of systems (H$_2$ + H$^+$, HD + H$^+$, SH$^+$ + H and CH$^+$ + H) of interest in astrochemistry and astrophysics.
		For all of the systems considered, an error of less than a factor 2 was found, at least for the dominant transitions and at low temperatures, which is sufficiently accurate for applications in the above mentioned disciplines.
	\end{abstract}
	
	\maketitle
	
	\section{Introduction}

	The computation of state-to-state (quantum state resolved) scattering cross sections and rate coefficients for collisional processes from first principles is one of the central objectives of theoretical chemical dynamics. These quantities are of interest in their own right, but also in various disciplines, such as astrochemistry and atmospheric chemistry, where they can be used in the interpretation of measured molecular spectra when the energy levels are not populated according to the Boltzmann distribution law (i.e., when the medium is not at local thermodynamic equilibrium), or in chemical networks\cite{Faure2020}. 
	
	It is now routinely possible to compute state-to-state cross sections and rate coefficients for (non-)reactive gas-phase collisions involving a low number of internal degrees of freedom with highly accurate fully quantum mechanical methods\cite{Zhang,Althorpe,Honvault_rev} (e.g., the time-independent rigid-rotor close-coupling (CC) method and the time-dependent wave packet (TDWP) approach). However, the application of the aformentioned quantum methods is considerably complicated when the collision proceeds via the formation and decay of an intermediate collision complex, because of the presence of an energy well on the corresponding electronic potential energy surface (PES)\cite{Gonzalez-Lezana20017}, requiring the inclusion of many closed channels in order to reach convergence in case of the CC method\cite{Loreau_2018}. In case of the TDWP method, a deep  potential well necessitates the use of dense grids and long propagation times, particularly at low collision energies. For such cases, statistical theories of scattering have been shown to provide reasonably accurate approximate results. This comes as no surprise given the statistical nature of complex-forming collisions, because of the long lifetime of such a collision complex\cite{Gonzalez-Lezana20017,DF9674400014}. Among the first to apply a statistical theory of scattering to, in particular, chemical reactions were Pechukas and Light, Bernstein and Miller \cite{Bernstein1963,PechukasLight1964,PechukasLight1965,Miller1970}. 
	More recently, Larr\'{e}garay \textit{et al.}\cite{Larregaray} introduced their mean potential phase space theory (MPPST) for chemical reactions, based on earlier work on phase space theory (PST)\cite{Bonnet,B700906B}.
	The older, but more sophisticated statistical theory of atom-diatom insertion reactions by Rackham \textit{et al}.\cite{Rackham2001,Rackham2003}, obtained by combining the statistical view with the coupled-channel capture theory of Clary and Henshaw\cite{Clary}, seems a popular choice for the computation of cross sections and rate coefficients for systems that behave statistically. It was later extended to include diatom-diatom collisions by Dagdigian\cite{Dagdigian}. While their statistical quantum method is capable of predicting rather accurate integral scattering cross sections (ICS), it is still computationally demanding to apply since it requires solving the close-coupled equations, albeit not for the entirety of the molecular configuration space. Another statistical approach that seems to perform well for reactive processes, provided the temperature is not too low, is the so-called statistical quasi-classical trajectory (SQCT) method\cite{Aoiz,Aoiz2,Jambrina}.
	
	However, in the present article the focus is on the simpler statistical adiabatic channel model (SACM), as conceptualized by Quack and Troe in the mid 1970's \cite{Quack&Troe2,Quack&Troe3,Quack&Troe_inform}. Their approach to the calculation of state-to-state integral scattering cross sections (ICS) is fundamentally statistical, based on the work by Pechukas, Light and Miller, but the way in which they count the number of accessible entrance and exit channels is quite different from the  statistical theories mentioned above. To this end, ro-vibronic adiabatic potential curves (\textit{adiabats}, for short) were used based on the long-range potential. These are radial potentials obtained by correlating the different modes of motion of the collision complex to the modes of motion of its dissociation products. Furthermore, the statistical adiabatic channel model falls under the category of adiabatic capture theories\cite{Clary_ad,Smith}.
	Recently, the interest in this SACM approach was renewed with the introduction of an improved way of constructing the adiabatic potential curves \cite{Loreau_2018}, as described in section \ref{sec.2B}. The authors showed, using this new way of computing adiabatic curves in their SACM-inspired approach, that the ro-vibrationally resolved rate coefficients for rotationally inelastic collisions for a number of benchmark systems were in excellent agreement with accurate rigid-rotor close-coupling results, at least when the systems were characterized by a deep potential energy well on their electronic potential energy surface (PES). When such a well is present, an error of less then a factor 2-3 was found for the dominant transitions at low temperature, with the error decreasing for increasing well depth of the intermediate complex and for decreasing temperatures\cite{Loreau_2018,Faure2020,Loreau2018}. Indeed, we should emphasize that this approach is capable of reproducing rather accurate state-to-state rate coefficients down to very low temperatures which, together with its computational efficiency, is one of its greatest strengths. 
	The necessity of a deep well on the PES and low temperatures in order for the SACM to produce accurate results was rationalized based on the general applicability constraints of the statistical approach, namely the lifetime of the collision complex, and the possibility of randomization of the available energy over the different internal degrees of freedom of this complex.
	
	In this paper, we report results obtained from an extension of the benchmarking process of the SACM-inspired approach, to include complex-mode atom-diatom collisions where there is competition between inelastic and reactive processes.
	This article is structured as follows: Section \ref{sec.2} is devoted to a discussion of the statistical approach to complex-forming atom-diatom collisions in the current implementation of the statistical adiabatic channel model. In section \ref{sec.3}, the results obtained with this model are then compared to accurate rate coefficients from the literature, and this for four benchmark systems: H$_3^+$, DH$_2^+$, SH$_2^+$ and CH$_2^+$. 
	Finally, in section \ref{sec.4} we summarize our findings and attempt to provide an outlook regarding future research on the topic.
	
	\section{Theoretical Methods}\label{sec.2}
	
	In the statistical model of molecular collision theory, an electronically adiabatic complex-forming atom-diatom collision of the type,
	\begin{eqnarray}\nonumber
		\mathrm{A} + \mathrm{BC}\,(v,j) \longrightarrow \left\{
		\begin{array}{l}
		\mathrm{A} + \mathrm{BC}\,(v',j')	\\
		\mathrm{C} + \mathrm{AB}\,(v',j')	\\
		\mathrm{B} + \mathrm{AC}\,(v',j'),
		\end{array}
	\right.
	\end{eqnarray}
	is viewed as a series of 2 consecutive, independent events\cite{Miller1970,Quack&Troe3}: 
	\begin{enumerate}
	\item 
	The collision partners come together in the entrance channel, $
	 \{E,J,M_J,\Pi,\alpha,v,j,l\}$, to form the collision complex, at which point the history of formation of the complex is forgotten; that is, with the exception of the constants of motion characterized by the set of good quantum numbers, $\{ E,J,M_J,\Pi \}$.
	\begin{eqnarray}\nonumber
		\mathrm{A} + \mathrm{BC}\,(v,j) \longrightarrow [\mathrm{ABC}].
	\end{eqnarray}
	The probability of this event is referred to as the capture probability, $p_{\alpha vjl}^{JM_J\Pi}(E)$.
	\item 
	The sufficiently long-lived complex decomposes into any of the exit channels, $\{E,J,M_J,\Pi,\alpha',v',j',l' \}$, that are accessible for the aforementioned set of good quantum numbers, and this with probability, $p_{\alpha' v'j'l'}^{JM_J\Pi}(E)/\left(\sum_{\alpha'' v''j''l''}p_{\alpha'' v''j''l''}^{J M_J \Pi}(E) \right)$.
	\begin{eqnarray}\nonumber
		[\mathrm{ABC}] \longrightarrow \left\{
		\begin{array}{l}
			\mathrm{A} + \mathrm{BC}\,(v',j')	\\
			\mathrm{C} + \mathrm{AB}\,(v',j')	\\
			\mathrm{B} + \mathrm{AC}\,(v',j').
		\end{array}
		\right.
	\end{eqnarray}
	\end{enumerate}  
	
	$E$ is the total energy of the collisional system and $J$ is the total angular momentum quantum number, wich is related to the rotational quantum number of the diatomic, $j$ (or $j'$), and the orbital angular momentum quantum number, $l$ (or $l'$), by the usual relation arising from angular momentum coupling ($\hat{\vb{J}} = \hat{\vb{j}} + \hat{\vb{L}}$): $J = \abs{j-l}, \dots, j+l$ (or $J = \abs{j'-l'}, \dots, j'+l'$). $M_J$ is the quantum number controlling the projection of the total angular momentum and $\Pi$ is the triatomic inversion parity quantum number. 
	$\alpha$ and $\alpha'$ are indices indicating the arrangement, $\alpha$ being the entrance arrangement ($\mathrm{A} + \mathrm{BC}$) and $\alpha'$ the exit arrangement ($\mathrm{A} + \mathrm{BC}$, $\mathrm{C} + \mathrm{AB}$ or $\mathrm{B} + \mathrm{AC}$), and thus whether the collision is reactive ($\alpha' \neq \alpha$) or non-reactive ($\alpha' = \alpha$).  
	
	The probability for the transition $(\alpha,v,j,l) \rightarrow (\alpha',v',j',l')$, given by the square modulus of the corresponding $\vb{S}$-matrix element, is then expressed as,
	
	\begin{equation}\label{Eq.2.1}
		\abs{S^{J M_J \Pi}_{\alpha' v'j'l',\alpha vjl}(E)}^2 = \dfrac{p_{\alpha vjl}^{JM_J\Pi}(E) p_{\alpha' v'j'l'}^{JM_J\Pi}(E)}{\sum_{\alpha'' v''j''l''}p_{\alpha'' v''j''l''}^{J M_J \Pi}(E)} \equiv P_{\alpha' v'j'l',\alpha vjl}^{J M_J \Pi}(E).
	\end{equation}
    The possibility of direct transitions (due to an abstraction mechanism) has been neglected in Eq. (\ref{Eq.2.1}), thereby limiting the applicability of this equation to complex-mode processes exclusively (i.e. processes were a collision complex is formed in the course of the process), which is our sole interest here.
	Based on conservation of probability, one can show that Eq. (\ref{Eq.2.1}) can be rewritten as,
	\begin{equation}\label{Eq.2.2}
		 P_{\alpha' v'j'l',\alpha vjl}^{J M_J \Pi}(E) = \left\{ 
		\begin{array}{ll}
		\quad \hspace{17pt} 0 & N(E,J,M_J,\Pi) = 0\\	
		\dfrac{1}{N(E,J,M_J,\Pi)} & N(E,J,M_J,\Pi) \neq 0,
		\end{array}
		\right.
	\end{equation}
	where $N(E,J,M_J,\Pi)$ is the total number of open exit channels at specified values of the good quantum numbers $E,J,M_J$ and $\Pi$. 
	A criterion on the basis of which channels are classified as open or closed is given at the end of Section II C. 
	Eq. (\ref{Eq.2.2}) expresses that dissociation of the complex in the open exit channels happens with equal probability.
	
	\subsection{Statistical cross sections}\label{sec.2A}
	
	The state-to-state integral scattering cross sections (ICS) for an atom-diatom collision are given by following 
	expression from quantum scattering theory\cite{QuackSymm},
	\begin{eqnarray}
		\sigma_{\alpha' v'j',\alpha vj} (E)=&&\dfrac{\pi \hbar^2}{2\mu_R E_c(2j+1)} \sum_{J=0}^{+\infty} \sum_{M_J=-J}^{+J} \sum_{\Pi =-1}^{+1}\nonumber \\ \label{eq.2.A2}  
		\times \sum_{l=\abs{J-j}}^{J+j} \sum_{l'=\abs{J-j'}}^{J+j'} &&\abs{\delta_{\alpha' \alpha} \delta_{v'v} \delta_{j'j}\delta_{l'l}-S^{J M_J \Pi}_{\alpha' v'j'l',\alpha vjl}(E)}^2,
	\end{eqnarray}
	where $E_{\alpha vj}$ is the energy of the ro-vibrational state of the diatomic molecule in the entrance channel, $E_c \equiv E-E_{\alpha vj}$ is the initial collision energy, and $\mu_R$ is the reduced mass of the atom-diatom system in the entrance arrangement.
	Since the focus in this article is on rotationally inelastic collisions ($\alpha'=\alpha$ and $j'\neq j$), vibrationally elastic or otherwise ($v'=v$ or $v'\neq v$ ), as well as rotational transitions in reactive collisions ($\alpha' \neq \alpha$), Eq. (\ref{eq.2.A2}) can be simplified,
	\begin{eqnarray}
		\sigma_{\alpha' v'j',\alpha vj} (E)=\dfrac{\pi \hbar^2}{2\mu_R E_c(2j+1)} &&\sum_{J=0}^{+\infty} \sum_{M_J=-J}^{+J} \sum_{\Pi =-1}^{+1} \nonumber \\ \times \sum_{l=\abs{J-j}}^{J+j} \sum_{l'=\abs{J-j'}}^{J+j'} &&\abs{S^{J M_J \Pi}_{\alpha' v'j'l',\alpha vjl}(E)}^2.
	\end{eqnarray}
	Use of Eq. (\ref{Eq.2.2}) results in,
	\begin{eqnarray}\label{Eq.2.A4}
		\sigma_{\alpha' v'j',\alpha vj} &&(E)= \dfrac{\pi \hbar^2}{2\mu_R E_c(2j+1)} \sum_{J=0}^{+\infty} (2J+1) \nonumber \\
		 \times &&\sum_{\Pi =-1}^{+1} \dfrac{N(E,J,\Pi,\alpha,v,j)N(E,J,\Pi,\alpha',v',j')}{N(E,J,\Pi)}, 
	\end{eqnarray}
	where $N(E,J,\Pi,\alpha,v,j)$ and $N(E,J,\Pi,\alpha',v',j')$ were obtained by summing over the orbital angular momentum quantum numbers, $l$ and $l'$ respectively, and $\sum_{M_J=-J}^{+J} \abs{S^{J M_J \Pi}_{\alpha' v'j',\alpha vj}(E)}^2 = (2J+1) \abs{S^{J \Pi}_{\alpha' v'j',\alpha vj}(E)}^2$.
	
	\subsection{Ortho-/Para-H$_2$ Separation and Identical Nuclei}
	
	The collisions of interest in this paper are of the form (charges are implied),
	\begin{eqnarray}\nonumber
		\mathrm{B'} + \mathrm{AB}\,(v,j) \longrightarrow \left\{
		\begin{array}{l}
			\mathrm{B'} + \mathrm{AB}\,(v',j') \quad (1)	\\
			\mathrm{B} + \mathrm{AB'}\,(v',j') \quad (2)\\
			\mathrm{A} + \mathrm{B_2}\,(v',j') \hspace{15pt} (3),
		\end{array}
		\right.
	\end{eqnarray}
	where 
	A $\equiv$ C, S, D, H and B $\equiv$ H
	, and typically involving some competition between inelastic (1), exchange (2) and insertion processes (3). 
	These collisions involve molecular hydrogen, which can occur as 2 different nuclear spin isomers: ortho- and para-H$_2$, with total nuclear spin quantum numbers, $I=1$ and $I=0$ respectively\cite{Lique2012,H3pSQM}. Because the nuclear wavefunction for the collisional system has to obey the proper exchange symmetry with respect to identical nuclei, $o$-H$_2$ can only occur with odd rotational quantum numbers, while $p$-H$_2$ is characterized by even values for $j$. The $o$-H$_2$ $\rightleftharpoons$ $p$-H$_2$ conversion is impossible via inelastic collisions. It is, however, possible via exchange with for example protons\cite{Lique2012}. In order to account for this type of "selection rule", it will prove to be important to compute statistical cross sections according to Eq. (\ref{Eq.2.A4}) for these benchmark systems, by performing a separation based on the nuclear spin quantum number of H$_2$. So the (total) number of open exit channels, $N(E,J,\Pi,I)$ and $N(E,J,\Pi,I,\alpha',v',j')$, and the number of open entrance channels, $N(E,J,\Pi,I,\alpha,v,j)$, and thus $P_{\alpha' v'j',\alpha vj}^{J\Pi I}(E)$, are determined, keeping track of the nuclear spin of H$_2$. This is valid if we assume a Hamiltonian that does not depend on nuclear spin, the latter consequently being a constant of motion\cite{QuackSymm}.
	In addition, a statistical weight of $\frac{1}{2}$ has to be applied to $P_{\alpha' v'j',\alpha vj}^{J\Pi I}(E)$. 
	
	\subsubsection{The case of $\mathbf{B = H \;\text{and}\; A \neq H}$}
	The scattering of $\mathrm{H'} + \mathrm{AH}$, where $\mathrm{A} \neq \mathrm{H}$, involves the formation of H$_2$ through the reactive channels (3). For the reverse reaction, 
		\begin{eqnarray}\nonumber
		\mathrm{A} + \mathrm{H_2}\,(v,j) \longrightarrow \left\{
		\begin{array}{l}
			\mathrm{A} + \mathrm{H_2}\,(v',j') \hspace{15pt} (1)	\\
			\mathrm{H} + \mathrm{AH'}\,(v',j') \quad (2)\\
			\mathrm{H'} + \mathrm{AH}\,(v',j') \hspace{10pt} (3),
		\end{array}
		\right.
	\end{eqnarray}
	 the above mentioned "selection rule" comes into play for the inelastic process (1), requiring separate calculations for ortho- and para-H$_2$. In order to be consistent with the microscopic reversibility property, and thus detailed balance, such a separation should also be perfomed for the actual scattering of interest. That means that the $\mathrm{H'} + \mathrm{AH}$ scattering is viewed as follows:
	 \begin{eqnarray}\nonumber
	 	\mathrm{H'} + \mathrm{AH}\,(v,j) \longrightarrow \left\{
	 	\begin{array}{l}
	 		\frac{1}{2}\left(\mathrm{H'} + \mathrm{AH}\,(v',j')\right) \quad\\
	 		\frac{1}{2}\left(\mathrm{H} + \mathrm{AH'}\,(v',j')\right) \quad\\
	 		\mathrm{A} + o\text{-}\mathrm{H_2}\,(v',j') \hspace{15pt}(j' \,\text{odd}) \\
	 		\mathrm{A} + p\text{-}\mathrm{H_2}\,(v',j') \hspace{15pt}(j' \,\text{even}).
	 	\end{array}
	 	\right.
	 \end{eqnarray}
	Statistical integral scattering cross sections for inelastic and exchange processes are then calculated as,
	\begin{eqnarray}
		\sigma^{\text{exch}}_{v'j',vj}(E)=\sigma^{\text{inel}}_{v'j',vj}(E)=\dfrac{\sigma^{I=0}_{v'j',vj}(E)+ \sigma^{I=1}_{v'j',vj}(E)}{2},
	\end{eqnarray}
	while for the insertion reaction,
	\begin{eqnarray}
		\sigma^{\text{inser}}_{v'j',vj}(E)=\sigma^{I=0}_{v'j'=\text{even},vj}(E)+\sigma^{I=1}_{v'j'=\text{odd},vj}(E).
	\end{eqnarray}
	Note that this is consistent with the assumption of distinguishable particles.
	
	\subsubsection{The case of $\mathbf{B = A = H}$}
	
	Also for this specific case, we apply the above mentioned procedure for the computation of the statistical integral cross sections. More concretely, this means that one views the scattering of H$_2$ by protons as follows:
	\begin{eqnarray}\nonumber
		\mathrm{H'^+} + o\text{-}\mathrm{H_2}\,(v,j) \longrightarrow \left\{
		\begin{array}{l}
			\mathrm{H'^+} + o\text{-}\mathrm{H_2}\,(v',j') \hspace{23pt} (j,j' \,\text{odd})\\
			\mathrm{H^+} + o\text{-}\mathrm{HH'}\,(v',j') \qquad (j,j' \,\text{odd}) \\
			\mathrm{H^+} + p\text{-}\mathrm{H'H}\,(v',j')  (j\,\text{odd}, j'\,\text{even}),
		\end{array}
		\right.
	\end{eqnarray}
	and a similar reasoning applies for para-H$_2$ in the entrance channel.
	In addition, statistical weights have to be applied, in order to account for the indistinguishability of the 3 identical protons. This is done through the procedure first outlined in Ref. \onlinecite{Miller1969} by Miller, and more recently by Grozdanov \textit{et al.}\cite{Grozdanov}, based on the work by Park and Light (see Ref. \onlinecite{Park&Light}). The transition probabilities for an inelastic process ($\alpha' = \alpha$), $P_{\alpha v'j',\alpha vj}^{J\Pi}(E)$, are expressed as,
    \begin{eqnarray} \label{Eq.gr_ine}
		P_{\alpha v'j',\alpha vj}^{J\Pi}(E) = \left\{
		\begin{array}{l}
			\frac{2}{9} P_{\alpha v'j',\alpha vj}^{J\Pi (1)} + \frac{2}{9} P_{\alpha v'j',\alpha vj}^{J\Pi (2)} \hspace{3pt} (j,j' \,\text{odd}) \\
			\frac{2}{3} P_{\alpha v'j',\alpha vj}^{J\Pi (2)} \hspace{52pt} (j, j'\,\text{even}) \\
			0 \hspace{60pt} \hspace{25pt}(\text{otherwise}),
		\end{array}
		\right.
	\end{eqnarray}
	while the transition probabilities for a reactive process ($\alpha' \neq \alpha$), $P_{\alpha' v'j',\alpha vj}^{J\Pi}(E)$, are given by:
	\begin{eqnarray} \label{Eq.gr_ex}
		P_{\alpha' v'j',\alpha vj}^{J\Pi}(E) = \left\{
		\begin{array}{l}
			\frac{4}{9} P_{\alpha' v'j',\alpha vj}^{J\Pi (1)} + \frac{1}{9} P_{\alpha' v'j',\alpha vj}^{J\Pi (2)} \hspace{7pt} (j,j' \,\text{odd})\\
			P_{\alpha' v'j',\alpha vj}^{J\Pi (2)} \hspace{48pt} (j\,\text{even}, j'\,\text{odd}) \\
			\frac{1}{3}P_{\alpha' v'j',\alpha vj}^{J\Pi (2)} \hspace{55pt} (\text{otherwise}).
		\end{array}
		\right.
	\end{eqnarray}
	Since the focus in this article will be on the sum of inelastic and reactive rate coefficients, the transition probabilities are a simple sum of Eqns. (\ref{Eq.gr_ine}) and (\ref{Eq.gr_ex}), $P_{v'j',vj}^{J\Pi}(E) = P_{\alpha' v'j',\alpha vj}^{J\Pi}(E) + P_{\alpha v'j',\alpha vj}^{J\Pi}(E)$:
	\begin{eqnarray}\label{Eq.gr_tot}
		P_{v'j',vj}^{J\Pi}(E) = \left\{
		\begin{array}{l}
			\frac{2}{3} P_{v'j',vj}^{J\Pi (1)}(E) + \frac{1}{3} P_{ v'j',vj}^{J\Pi (2)}(E) \hspace{16pt} (j,j' \,\text{odd})\\
			\frac{1}{3} P_{v'j',vj}^{J\Pi (2)}(E) \hspace{53pt} (j\,\text{odd}, j'\,\text{even}) \\
			P_{v'j',vj}^{J\Pi (2)}(E) \hspace{28pt} (j\, \text{even}, j'\,\text{odd or even}).
		\end{array}
		\right.
	\end{eqnarray}
    In the equations above, $P_{\alpha' v'j',\alpha vj}^{J\Pi (1)}(E)$ and $P_{\alpha' v'j',\alpha vj}^{J\Pi (2)}(E)$ differ in whether even rotational states of H$_2$ are counted as accessible exit channels or not: the superscript (2) indicates that both even and odd states are counted, corresponding to the $E$ irreducible representation of the $S_3$ nuclear permutation symmetry group, whereas the superscript (1) denotes that only odd states are considered (corresponding to the $A_2$ irreducible representation of $S_3$)\cite{Grozdanov,HonvaultP2011OHcb}.

	\subsection{Adiabatic Potential Curves}\label{sec.2B}

	In the statistical adiabatic channel model, a count of the open entrance/exit channels for a given arrangement, $N(E,J,\Pi,\alpha,v,j)$, $N(E,J,\Pi,\alpha',v',j')$ and $N(E,J,\Pi)$, is provided based on an analysis of ro-vibronic adiabatic potential curves, $V_{\alpha vjl}^{J\Pi}(R)$ and $V_{\alpha' vjl}^{J\Pi}(R)$. The way these curves were computed by Quack and Troe is described in Ref. \onlinecite{Quack&Troe2}.

	The new and improved way of computing \textit{adiabats} is outlined in Ref. \onlinecite{Loreau_2018} and differs from the original approach. It can be best understood by considering the coupled differential equations that appear in the time-independent rigid-rotor close-coupling (CC) method. The coupled equations are obtained by expanding the total nuclear wavefunctions, $\ket{\Psi}$, in a basis set of angular functions, ${\ket{\gamma}}$, the nature of which depends on the collision partners\cite{Flower2007}:
	\begin{equation}
	\ket{\Psi} = \dfrac{1}{R} \sum_{\gamma} \chi_\gamma(R) \ket{\gamma}.
	\end{equation}
	The expansion coefficients, $\chi_\gamma(R)$, depend on the radial distance, $R$, between the atom and the center of mass of the diatomic molecule, and can be determined after substitution in the time-independent nuclear Schr\"{o}dinger equation for relative motion with Hamiltonian in Jacobi coordinates,
	\begin{eqnarray}\label{Eq.2.B1.1}
		\hat{H} &&= -\dfrac{\hbar^2}{2\mu_R} \pdv[2]{}{R} + \dfrac{(\hat{\vb{J}}-\hat{\vb{j}})^2}{2 \mu_R R^2} + V(R,\varphi;r) + \hat{h} \\ \label{Eq.2.B1.2}
		\hat{h} && = -\dfrac{\hbar^2}{2\mu_r} \pdv[2]{}{r} + \dfrac{\hat{\vb{j}}^2}{2 \mu_r r^2}.
	\end{eqnarray}
	In Eq. (\ref{Eq.2.B1.1}), $(\hat{\vb{J}}-\hat{\vb{j}})^2$ is the square of the orbital angular momentum operator, $\hat{h}$ is the contribution arising from ro-vibrational motion of the diatomic molecule, and $V(R,\varphi;r)$ is the electronic PES in Jacobi coordinates for a fixed value of the internuclear distance of the diatomic, $r$ ($\varphi$ is the angle between the radial and diatomic vectors). 
	Integration over the angles on which $\ket{\gamma}$ depends leads to the coupled equations:
	\begin{eqnarray}\label{Eq.2.B2}
	-\dfrac{\hbar^2}{2\mu_R} \pdv[2]{\chi_\gamma}{R} + 
	\sum_{\gamma'} \mel{\gamma'}{\dfrac{(\hat{\vb{J}}-\hat{\vb{j}})^2}{2 \mu_R R^2}+\hat{h}+V}{\gamma}\chi_\gamma = E\chi_\gamma.
	\end{eqnarray}
	The \textit{adiabats} are computed by diagonalization of the Hamiltonian, excluding the radial kinetic energy contribution in Eq. (\ref{Eq.2.B2}). In other words, by diagonalization of the matrix representative of the operator,
	\begin{equation}\label{Eq.2.B3}
		\dfrac{(\hat{\vb{J}}-\hat{\vb{j}})^2}{2 \mu_R R^2}+\hat{h}+V(R,\varphi;r), 
	\end{equation}	
	in the basis of angular funtions, $\ket{\gamma}$, and this for different values of $J$ and $\Pi$. This can be done with available programs such as Molscat\cite{HutsonJeremyM2019mApf}, or with the recently developed in-house code. We should note that in the current way of computing the eigenvalues of Eq. (\ref{Eq.2.B3}), the diatomic molecule is considered as a rigid rotor (hence, why the dependence of $V$ on the diatomic internuclear distance, $r$, is parametric), so the only term in $\hat{h}$ that survives is the rotational contribution,
	\begin{eqnarray}
		\hat{h} = \dfrac{\hat{\vb{j}}^2}{2\mu_r r^2}.
	\end{eqnarray} 
	Consequently, only couplings between rotational states are taken into account, and vibration is not explicitly considered. It should also be noted that if the atom-diatom collision of interest is reactive, at least 2 different arrangements have to be considered and \textit{adiabats} have to be computed for these arrangements separately; that is, eigenvalues of Eq. (\ref{Eq.2.B3}) have to be computed for all relevant arrangements. The result is a set of adiabatic curves corresponding to the reactant arrangement and a set for each of the product arrangements.
	
	Asymptotically, for a given electronic state, the adiabatic potential curves can be associated with a rotational state of the diatomic molecule,
	\begin{eqnarray}\nonumber
		\lim\limits_{R_{\kappa} \rightarrow + \infty} V_{\kappa vjl}^{J\Pi}(R_{\kappa}) =&& V(r_\kappa) + B_{v\kappa} j(j+1)+ D_{v\kappa} j^2 (j+1)^2 \\ \label{Eq.2.B5}
	    &&+ H_\kappa j^3 (j+1)^3,
	\end{eqnarray}
    where $\kappa = \alpha$ or $\alpha'$. 
	In Equation (\ref{Eq.2.B5}), $V(r_{\kappa})$ is the constant electronic energy of the rigid diatom in the reactant or product arrangement, $B_{v\kappa}$  is the rotational constant for a given vibrational state, and the effect of centrifugal distortion (with $D_{v\kappa}$ and $H_{\kappa}$ the corresponding constants) has been accounted for up to third order. If the collision is reactive ($\alpha' \neq \alpha$), then $V(r_{\alpha'})-V(r_{\alpha})$ is the exothermicity or endothermicity of the reaction. In the current implementation, vibration is treated approximately by simply shifting the adiabatic curves by the difference in vibrational energy, and by fixing the internuclear distance of the diatomic molecule to its vibrationally averaged value. In this way, interactions between rotational states associated with different vibrational states are not accounted for, but since $\Delta E_{\mathrm{vibr}} \gg \Delta E_{\mathrm{rot}}$, this is not expected to introduce a large error.
	
	Counting channels is now done based on conservation of energy. If the total energy is greater than a centrifugal barrier, $\max_{R_{\kappa}} \{V_{\kappa vjl}^{J\Pi}\}$, that may exist in the entrance/exit channel corresponding to some arrangement, then that channel is considered open, and vice versa. If no barrier is present, then trivially:
	\begin{eqnarray}
		\max_{R_{\kappa}} \{V_{\kappa vjl}^{J\Pi}\} &&= \lim\limits_{R_{\kappa} \rightarrow + \infty} V_{\kappa vjl}^{J\Pi}(R_{\kappa}), 
	\end{eqnarray}
    so that asymptotically closed channels are considered closed. 
	Based on this simple notion, one can determine the number of open entrance and exit channels\cite{QuackSymm}:
	\begin{eqnarray}
		N(E,J,\Pi,\alpha,v,j) &&= \sum_{l}h(E - \max_{R_{\alpha}} \{V_{\alpha vjl}^{J\Pi}\}) \nonumber \\
		N(E,J,\Pi,\alpha',v',j') &&= \sum_{l'}h(E - \max_{R_{\alpha'}} \{V_{\alpha' v'j'l'}^{J\Pi}\}) \nonumber \\
		N(E,J,\Pi) &&= \sum_{\alpha'' v''j''} N(E,J,\Pi,\alpha'',v'',j'') \nonumber \\
		&&=  \sum_{\alpha'' v''j''l''} h(E - \max_{R_{\alpha''}} \{V_{\alpha'' v''j''l''}^{J\Pi}\}),
	\end{eqnarray}
	where $h$ is a unit Heaviside step function.
	Note that the criterion on the basis of which is decided whether a channel is open or closed is purely classical. No quantum tunneling through these \textit{adiabats} was considered in the current implementation of the SACM-inspired approach. It was observed by Dashevskaya \textit{et al.}\cite{Dashevskaya} that the neglect of tunneling in the adiabatic channel model did not cause large deviations for systems with a long-range attractive potential, provided that the temperatures are not extremely low (below $0.001$ K).
	
	\subsection{Rate Coefficients}\label{sec.2C}
		
	Once the state-to-state integral scattering cross sections are obtained by the procedures discussed in the previous subsections, temperature-dependent quantum-state (ro-vibrationally) resolved rate coefficients, $k_{\alpha' v'j',\alpha vj}(T)$, are computed in the standard way, assuming a Maxwell-Boltzmann distribution of collision energies,
	\begin{eqnarray}
		k_{\alpha' v'j',\alpha vj}(T) =&&
		\left( \dfrac{8}{\pi \mu_Rk_B^3 T^3} \right)^{1/2} \nonumber \\
		\times &&\int_{0}^{+ \infty} E_c\,e^{-E_{c}/k_B T}\,\sigma'_{\alpha' v'j',\alpha vj}(E_c)\,dE_c,
	\end{eqnarray}		
	wherre $\sigma'_{\alpha' v'j',\alpha vj}(E_c)$ is a cross section that depends on the collision energy.
	\section{Results and Discussion}\label{sec.3}
 	
 	In the following subsections we discuss the performance of the statistical adiabatic channel model (SACM) for atom-diatom insertion reactions, based on the results obtained for a number of systems. These benchmark systems were selected based on the following conditions.
 	\begin{enumerate}
 		\item[(i)] 
 		The presence of the desirable feature of a (deep) potential energy well on the electronic PES, such that statistical behavior can be expected.
 		\item[(ii)]
 		The availability of an accurate \textit{ab initio} global electronic PES for the electronic state of interest (typically the ground electronic state for the cases considered in this study). This is necessary for the construction of the adiabatic potential curves as discussed in \ref{sec.2B}.
 		\item[(iii)]
 		The availability of (state-to-state) rate coefficients as computed using accurate approaches, based on the same electronic PES used in (ii).
 	\end{enumerate}
 	
 	In order to make the comparison between SACM and high-level-of-theory (HLOT) results for benchmarking more quantitative, for each system, the weighted mean error factor (WMEF) was calculated for the temperatures considered. The WMEF is calculated as\cite{Loreau_2018},
 	\begin{eqnarray}
 		\mathrm{WMEF}\,(T) &&= \dfrac{\sum_{i} k_{i}^{\mathrm{HLOT}}r_i}{\sum_{i}k_{i}^{\mathrm{HLOT}}} \\
 		r_{i} &&= \max \left( \dfrac{k_{i}^{\mathrm{HLOT}}}{k_{i}^{\mathrm{SACM}}}, \dfrac{k_{i}^{\mathrm{SACM}}}{k_{i}^{\mathrm{HLOT}}} \right).
 	\end{eqnarray}
 	In this way, an error factor is defined such that $r_i \ge 1$. They are weighted by the high-level-of-theory rates, such that the dominant transitions, which are the ones of most interest, make the largest contributions to the mean. 
 	
 		\subsection{H$_{3}^{+}$-System}\label{3.D}
\begin{figure}
 		\centering
 		\includegraphics[width=1.1\columnwidth]{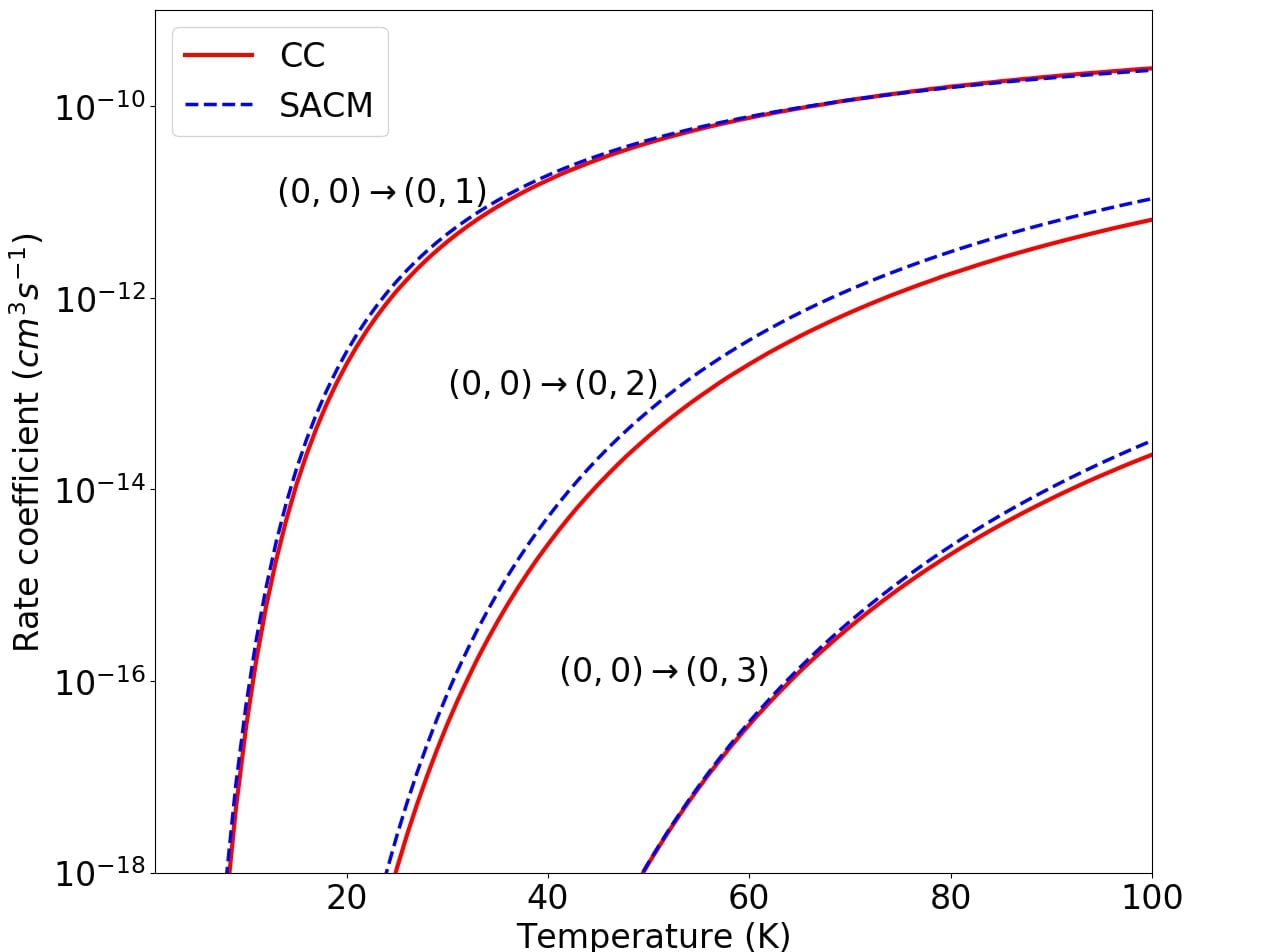}
 		\qquad
 		\includegraphics[width=1.1\columnwidth]{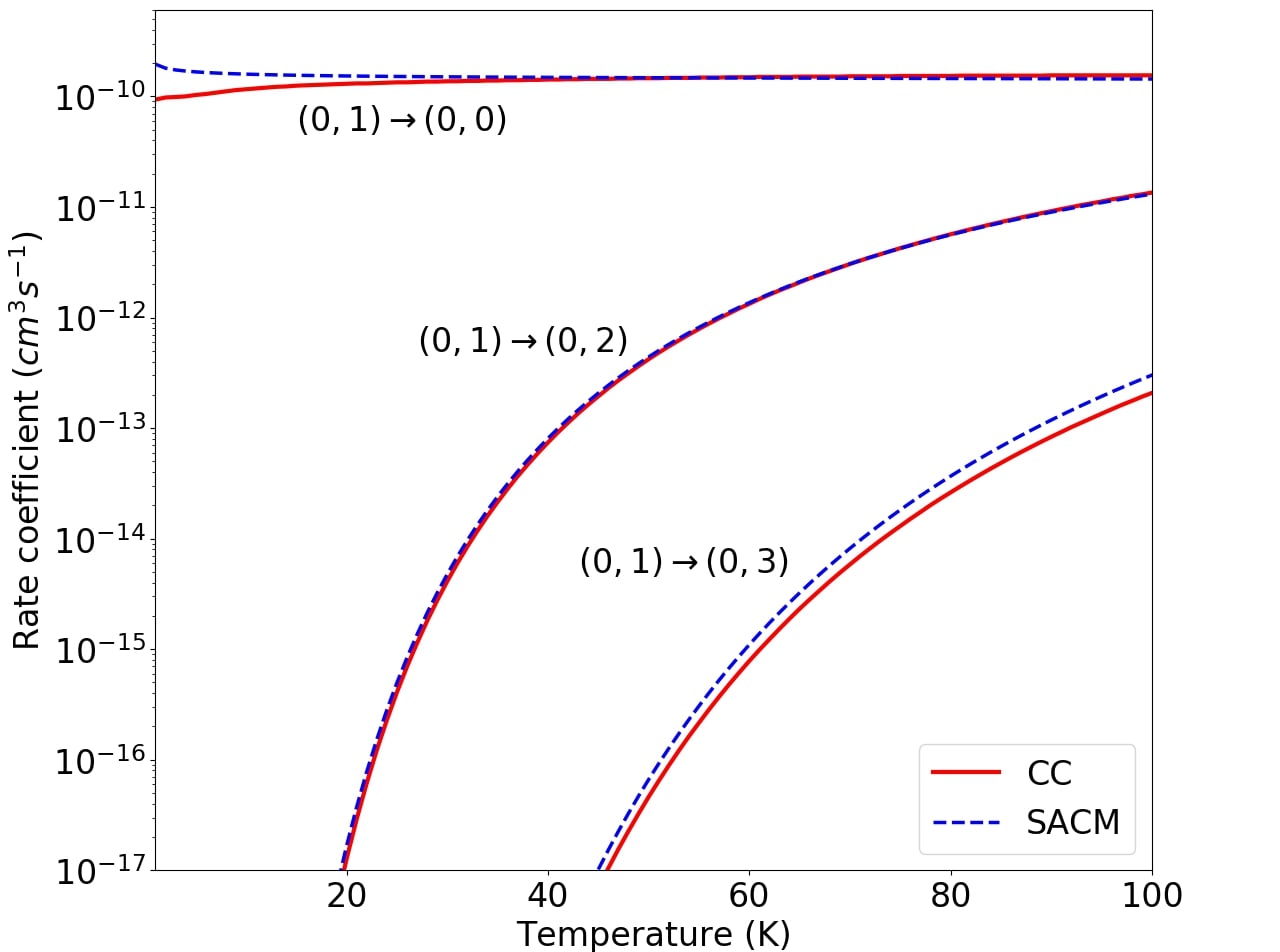}
 		\caption{Direct comparison of the CC and SACM rate coefficients as a function of temperature, for the $j=0 \rightarrow j'=1,2,3$ transitions (top panel) and the $j=1 \rightarrow j'=0,2,3$ transitions (bottom panel) in H$_2$, both by combined inelastic and reactive collisions. All these transitions are vibrationally adiabatic, where $v=v'=0$. The CC results were calculated by Honvault \textit{et al}. and Gonz\'{a}lez-Lezana \textit{et al}. (Refs. \onlinecite{Gonzalez-LezanaT2017,HonvaultP2011OHcb,Erratum_Honv}).}
 		\label{fig:h2hpj0jf123}
 	\end{figure}

\begin{figure}
 		\centering
 		\includegraphics[width=1.1\columnwidth]{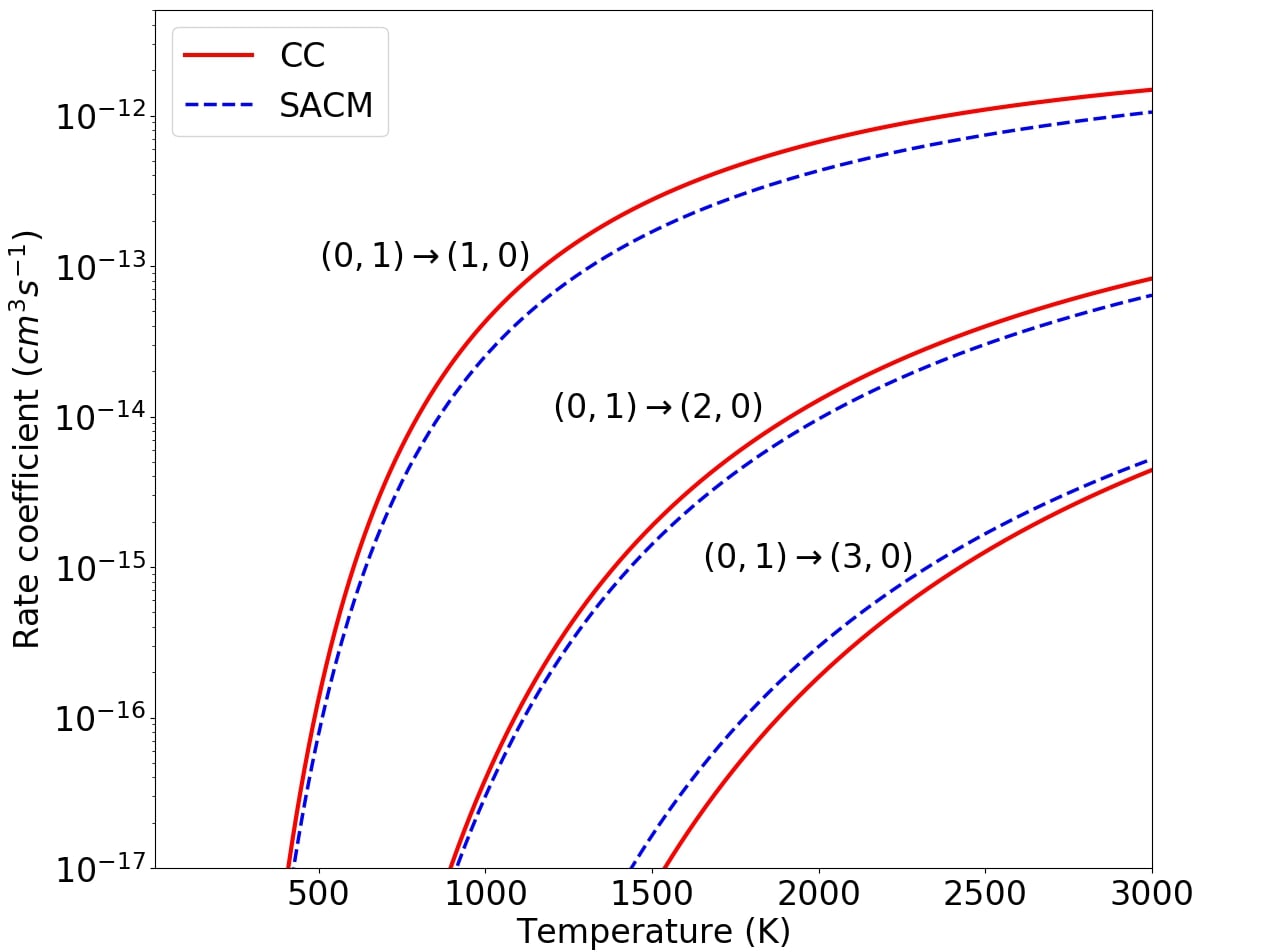}
 		\caption{Direct comparison of the CC and SACM rate coefficients as a function of temperature for the vibrational excitations, $v=0 \rightarrow v'=1,2,3$, and rotational deexcitation, $j=1 \rightarrow j'=0$, in H$_2$, both by combined inelastic and reactive collisions. The CC results were calculated by Gonz\'{a}lez-Lezana \textit{et al}. (Ref. \onlinecite{Gonzalez-LezanaT2017}).}
 		\label{fig:v0j1vfjf0}
 	\end{figure}

 	Adiabatic curves were computed based on the full-dimensional global VLABP PES of Velilla \textit{et al.} (see Ref. \onlinecite{Velilla}). This adiabatic electronic surface, which is invariant under nuclear permutations, was obtained by performing electronic structure calculations at the full configuration interaction (FCI) level of theory, using the aug-cc-pV6Z one-electron basis set. The ground electronic state, $\mathrm{H_3^+}(1^1A')$, has a global minimum of $\sim 4.6$ eV as measured with respect to the $\mathrm{H^+} + \mathrm{H_2}\,(X^{1}\Sigma^{+}_{g})$ asymptote. The formation of the H$_3^+$ complex is furthermore completely barrier-less. The competition between inelastic (A1) and exchange processes (A2 and A3), both involving H$_2$ ($X^{1}\Sigma^{+}_{g}$) and H$^+$, was considered in the SACM calculations.
 	\begin{eqnarray}\nonumber
 	\mathrm{H_{2}}\;(j,v) + \mathrm{H'^{+}}
 	&&\longrightarrow \mathrm{H_{2}}\;(j',v') + \mathrm{H'^{+}} \hspace{13pt} (\text{A}1)\\ \nonumber
 	\mathrm{HH}\;(j,v) + \mathrm{H'^{+}} && \longrightarrow \mathrm{HH'}\;(j',v') + \mathrm{H^{+}} \hspace{10pt} (\text{A}2)\\ \nonumber
 	\mathrm{HH}\;(j,v) + \mathrm{H'^{+}} && \longrightarrow \mathrm{H'H}\;(j',v') + \mathrm{H^{+}} \hspace{10pt} (\text{A}3)
 	\end{eqnarray}
 	The theoretical treatment of the H$_2$ + H$^+$ reaction and inelastic collision has been the focus of much attention (see for example Refs. \onlinecite{Gerlich,Lezana}).
 	Time-independent quantum mechanical (TIQM) calculations of ro-vibrationally resolved integral scattering cross sections were performed by Honvault \textit{et al}. and Gonz\'{a}lez-Lezana \textit{et al}. (see Refs. \onlinecite{Gonzalez-LezanaT2017,HonvaultP2011OHcb,Erratum_Honv}). The corresponding state-to-state rate coefficients, $k_{v'j',vj}(T)$, where $v=0$, $v'=\{0,1,2,3\}$ and $j=j'=\{0,1,2,3\}$, were used for benchmarking.  
 	It is important to note that in the quantum mechanical treatment of this system, inelastic and reactive processes cannot be distinguished\cite{HonvaultP2011OHcb}. Therefore, the SACM rate coefficients for both type of processes were summed.
 	Figure \ref{fig:h2hpj0jf123} shows a comparison of the SACM and CC rate coefficients over a limited temperature interval for purely rotational transitions with $v=v'=0$. Overall, the agreement is excellent, with SACM rate coefficients differing by a factor of less than 2 as compared to the close-coupling results.
 	In Figure \ref{fig:v0j1vfjf0} the focus is on the performance of the statistical adiabatic channel model for vibrational transitions. Vibrational transitions starting from the ground vibrational state of H$_2$ were considered, each time for the same ortho-para rotational transition ($j=1 \rightarrow j'=0$). With the exception of the $v=0 \rightarrow v'=3$ transition, again, we find good agreement between our SACM rate coefficients and the CC results, characterized by an error factor of less than 2. For the $v'=3$ case, such an error factor is only found at temperatures above $\sim$ 1600 K, where the rate coefficient becomes larger than $10^{-17}$cm$^{3}$s$^{-1}$.
 	

 	\subsection{DH$_{2}^{+}$-System}\label{3.C}
 	The 3-dimensional electronic PES for this system is the same as the one for the H$_3^+$ system (see Ref. \onlinecite{Velilla}), at least in the Born-Oppenheimer approximation neglecting non-adiabatic corrections that depend on mass. Consequently, within this approximation, the same global minimum is found. 
 	All species involved, HD ($X^{1}\Sigma^{+}$) and H$_2$ ($X^{1}\Sigma^{+}_{g}$), are in their electronic ground states, and competition between processes B1, B2 and B3 was accounted for in the statistical calculations. In contrast to the $\mathrm{H_3^+}$ isotopic variant discussed in \ref{3.D}, the insertion reaction for this system is slightly endothermic by $\sim 39.5$ meV due to the difference in vibrational zero-point energies of $\mathrm{H_2}$ and $\mathrm{HD}$, and the different ionisation energies of H and D \cite{LepersMaxence2019Qmso,10.1093/mnras/staa3146}.
 	\begin{eqnarray}\nonumber
 		\mathrm{HD}\;(j,v=0) + \mathrm{H'^{+}}
 		&&\longrightarrow \mathrm{HD}\;(j',v'=0) + \mathrm{H'^{+}} \hspace{13pt} (\text{B}1) \\ \nonumber
 		\mathrm{HD}\;(j,v=0) + \mathrm{H'^{+}}
 		&&\longrightarrow \mathrm{H'D}\;(j',v'=0) + \mathrm{H^{+}} \hspace{13pt} (\text{B}2) \\ \nonumber
 	    \mathrm{HD}\;(j,v=0) + \mathrm{H^{+}}
 		&&\longrightarrow \mathrm{D^{+}} + \mathrm{H_{2}}\;(j',v') \hspace{36pt} (\text{B}3)
 	\end{eqnarray}
 	Time-independent quantum scattering calculations for this system were recently performed by Desrousseaux \textit{et al.}\cite{Desrousseaux}, utilizing the close-coupling method. Accurate CC rate coefficients are available for rotational (de)excitation and for the reaction, and these results were used to investigate the performance of the SACM-inspired method for this system. As discussed in Ref. \onlinecite{Desrousseaux}, excellent agreement was found between the new close-coupling results and the SACM results for these processes, characterized by an error factor of less than 2. In fact, the largest differences reported were less than 65 $\%$.
 	Only purely rotational transitions were considered in that work however. Therefore, here we focus on the insertion reaction where H$_2$ can be ro-vibrationally excited. To this end, we used the rate coefficients, $k^{\text{inser}}_{v'j',00}$, as computed by Lepers \textit{et al}.\cite{LepersMaxence2019Qmso,10.1093/mnras/staa3146} with the TIQM method. 
 
\begin{figure}[H]
 		\centering
 		\includegraphics[width=1.1\columnwidth]{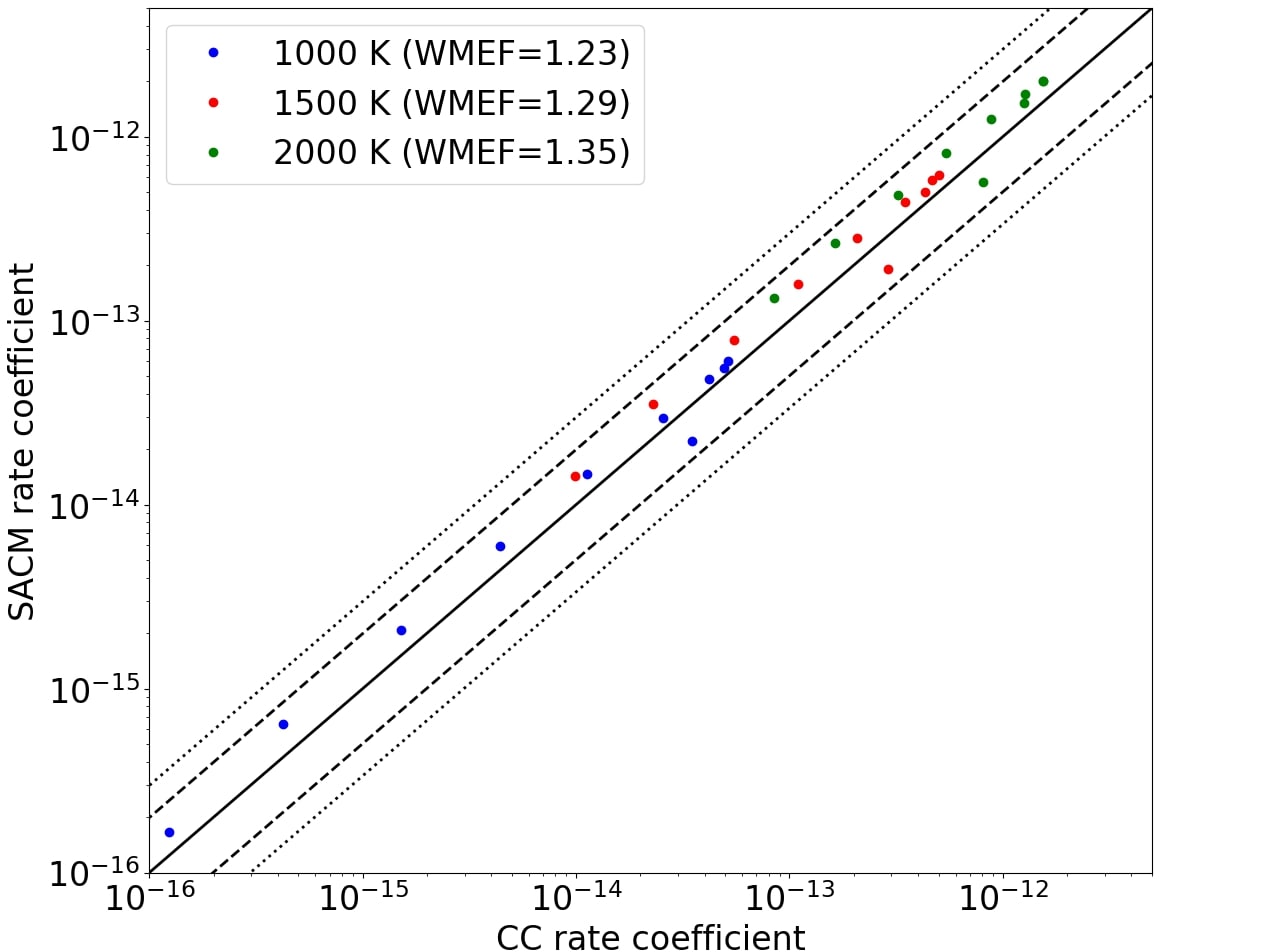}
 		\caption{Direct comparison at several temperatures of the SACM and the CC rate coefficients (in units of cm$^{3}$s$^{-1}$) for the insertion reaction, involving HD and H$^+$, for H$_2\,(v'=1)$. 
 		The dashed (dotted) lines represent an error factor of 2 (3). The CC results were computed by Lepers \textit{et al.} (Refs. \onlinecite{LepersMaxence2019Qmso,10.1093/mnras/staa3146}).}
 		\label{fig:v1hdhpscatter}
 	\end{figure}

 	Figure \ref{fig:v1hdhpscatter} shows the direct comparison of $k^{\text{inser}}_{v'j',00}$ for,  $v'=1$ and $j'=\{ 0,\dots,9 \}$, as computed with the SACM and CC methods. Overall, good agreement was found for most of the temperatures considered. An error factor of less than 2 was consistently found.
 	
 \newpage
 	\subsection{SH$_{2}^{+}$-System}\label{3.B}
	
 	For this system, the processes of interest are the inelastic collision (C1) and the exchange reaction (C2), both involving the sulphanylium cation, SH$^+$ ($X^{3}\Sigma^{-}$), and the hydrogen atom, H ($^{2}S$). The global full-dimensional electronic PESs for the quartet ($^4A''$) and doublet ($^2A''$) states of this system, both degenerate in the SH$^+$ ($X^{3}\Sigma^{-}$) + H ($^{2}S$) asymptote, were computed by Zanchet \textit{et al. }\cite{Zanchet_2019} using the internally contracted multireference configuration interaction (ic-MRCI) method, including singles and doubles, and accounting for Davidson correction (+Q). The aug-cc-pV5Z basis set was employed for all atoms. 
\begin{figure}[H]
   	 \centering
   	 \includegraphics[width=1.1\columnwidth]{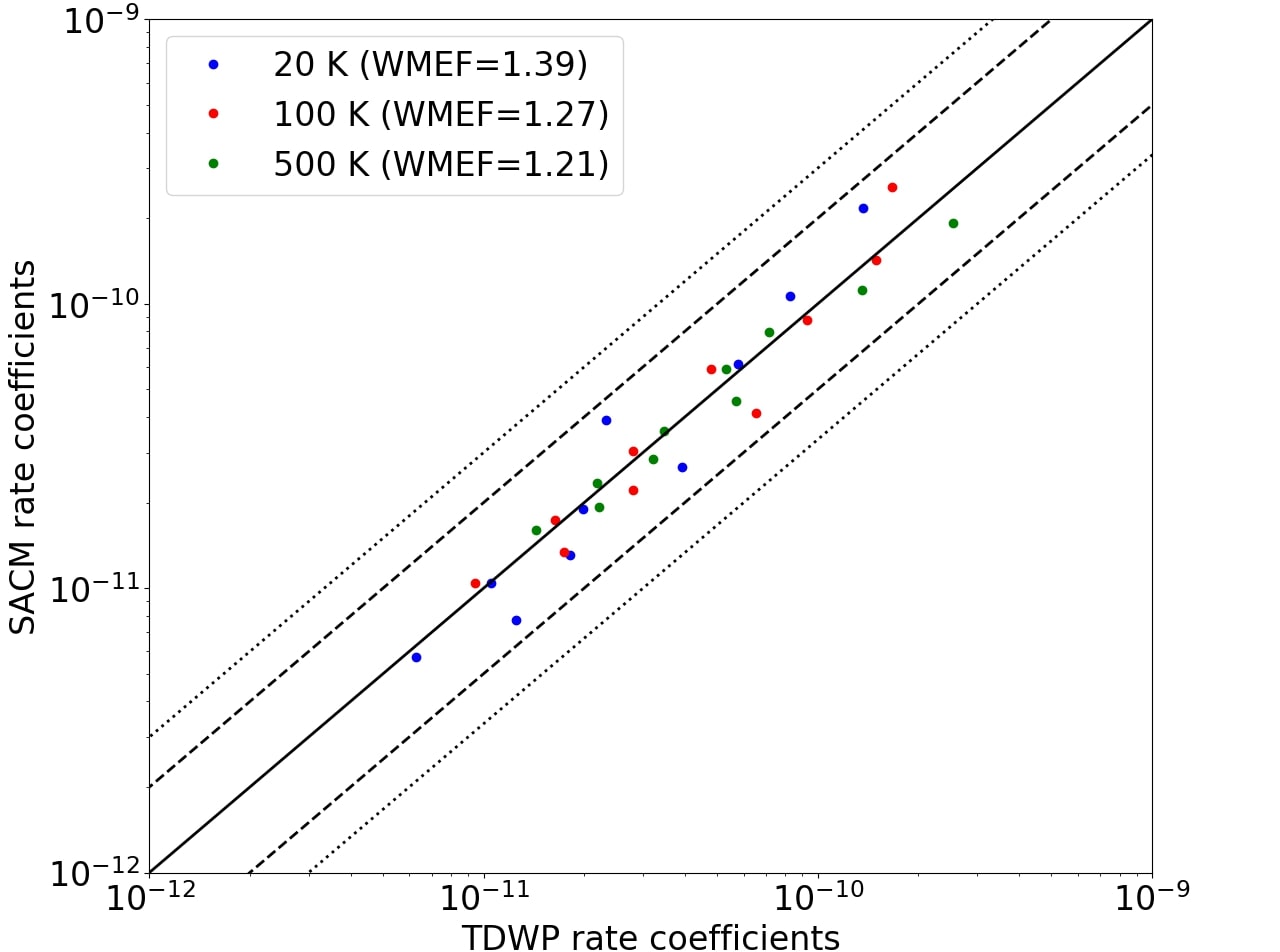}
   	 \qquad
   	 \includegraphics[width=1.1\columnwidth]{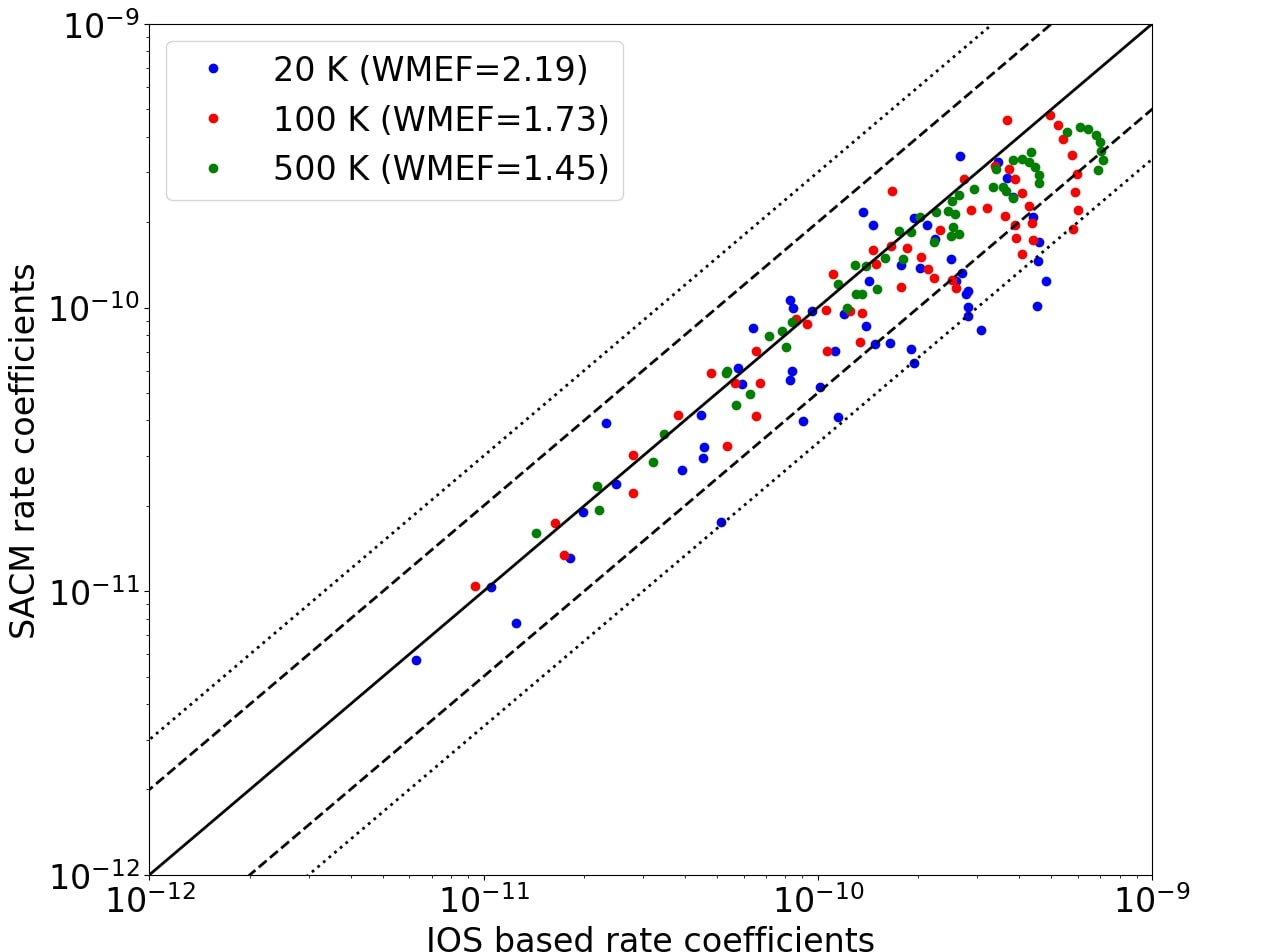}
   	 \caption{Direct comparison at several temperatures of the SACM and the TDWP (top panel) and IOS based (bottom panel) rate coefficients (in units of cm$^{3}$s$^{-1}$) for the rotational (de)excitation (inelastic collisions and exchange reactions), involving SH$^+$ and H. The dashed (dotted) lines represent an error factor of 2 (3). The TDWP and IOS based (methodology) results are presented in Refs. \onlinecite{Zanchet_2019,refId0,Faure&Lique_IOS}}.
   	 \label{fig:SH_Hp}
   \end{figure} 
	The quartet surface correlates to the S$^{+}$($^4S$) + H$_2$ ($X^{1}\Sigma^{+}_{g}$) asymptote, resulting in an exothermicity of $\sim 0.86$ eV, and shows almost no energy well\cite{Zanchet_2013}; dynamics on this surface is not considered in this study.
 	On the other hand, the doublet surface, correlating to the S$^{+}$($^2S$) + H$_2$ ($X^{1}\Sigma^{+}_{g}$) asymptote, predicts a potential energy well of $\sim 4$ eV and is without barrier for the formation of the intermediate complex. Because the insertion reaction on this doublet surface is endothermic by $\sim 1$ eV, no competition with the other processes is expected at low temperatures. Consequently, the reaction was not considered in the statistical quantum calculations provided that we restricted ourselves to low temperatures.
 	 \begin{eqnarray}\nonumber
 	  	\mathrm{SH^{+}}\;(j,v=0) + \mathrm{H'}
 	 	&&\longrightarrow \mathrm{SH^{+}}\;(j',v'=0) + \mathrm{H'} \hspace{13pt} (\text{C}1) \\ \nonumber
 	 	\mathrm{SH^{+}}\;(j,v=0) + \mathrm{H'}
 	 	&&\longrightarrow \mathrm{SH'^{+}}\;(j',v'=0) + \mathrm{H} \hspace{13pt} (\text{C}2)
 	  \end{eqnarray}
 	Scattering calculations combining the infinite-order-sudden (IOS) approximation and the time-dependent wavepacket (TDWP) method were presented in Refs. \onlinecite{Zanchet_2019,refId0}. Rotational deexcitation rate coefficients, $k^{\mathrm{exci}}_{0j',0j}(T)$, for process 1 and 2, summed, were used for benchmarking. The transitions for initial rotational states $j \in \{ 1,\dots, 10\}$ and final rotational state $j'=0$, were obtained with the time-dependent wave packet (TDWP) method. Results for all the other transitions, $j \in \{ 2,\dots, 10\}$ and $j'=j-1$, were computed with an IOS based method\cite{Faure&Lique_IOS}. Figure \ref{fig:SH_Hp} shows the direct comparison of the SACM results and the data. In particular, the upper panel shows a comparison with the TDWP results at temperatures of 20K, 100K and 500K. The lower panel does the same, only now including the IOS based results as well.
 	One can see that the agreement between SACM and quantum WP data, with the exception of some transitions, is rather good. The agreement with IOS based results is less good, especially at lower temperatures. This is expected since the accuracy of the IOS approximation decreases with decreasing temperature.

 	\subsection{CH$_{2}^{+}$-System}\label{3.A}	
 	
 	The full-dimensional electronic potential energy surface for this system, developed by Werfelli \textit{et al.} \cite{WerfelliGhofran2015Ltrc}, was used for the construction of the \textit{adiabats}. It was computed with the internally contracted multireference configuration interaction (ic-MRCI) method, accounting for Davidson correction (+Q), in conjunction with the aug-cc-pV5Z basis set.
 	The ground electronic state of the CH$_2^+$-system is characterized by a deep potential energy well of $\sim 4.8$ eV, which should favor a statistical treatment.
 	The interest is in the insertion reaction (D3) and inelastic collision (D1), both involving the methylidyne cation, CH$^+$($X^{1}\Sigma^{+}$), and the hydrogen atom, H ($^{2}S$), in their ground electronic states. Both scattering events are in competition with each other, as well as with an exchange reaction (D2); this competition was taken into account in the statistical calculations. Molecular hydrogen, H$_2$ ($X^{1}\Sigma^{+}_{g}$), and C$^+$($^{2}P$) are formed in their ground electronic states as products of the insertion reaction. This reaction is exothermic by $\sim 0.402$ eV.
 	\begin{eqnarray}\nonumber
 		\mathrm{CH^{+}}\;(j,v=0) + \mathrm{H'} \nonumber
 		&&\longrightarrow  \mathrm{CH^{+}}\;(j',v'=0) + \mathrm{H'} \hspace{13pt} (\text{D}1) \\\nonumber
 		\mathrm{CH^{+}}\;(j,v=0) + \mathrm{H'}
 		&&\longrightarrow  \mathrm{CH'^{+}}\;(j',v'=0) + \mathrm{H} \hspace{13pt} (\text{D}2) \\\nonumber
 		\mathrm{CH^{+}}\;(j,v=0) + \mathrm{H}
 		&&\longrightarrow  \mathrm{C^{+}} + \mathrm{H_{2}}\;(j',v') \hspace{36pt} (\text{D}3)
 	\end{eqnarray} 
 	This system has been the subject of earlier theoretical work\cite{Plasil_2011,Warmbier,GROZDANOV201323,Li,Bovino}, quantum or statistical, on the dynamics of CH$^+$. 
 	More recently, scattering calculations for this system were performed in Ref. \onlinecite{WerfelliGhofran2015Ltrc}, and these results were used for benchmarking.
 	Accurate rotational deexcitation rate coefficients for process 1, $k^{\mathrm{inel}}_{0j',0j}(T)$, are available for $j \in \{ 2,\dots,7 \}$ 
 	and $j' = j-1$. 
 	In addition, reaction rate coefficients for process 3, $k^{\mathrm{inser}}_{0j}(T)$, were computed for $j \in \{ 0,\dots,7 \}$. 
	Note that the reactive rate coefficients are only resolved up to the ro-vibrational state of CH$^+$, so they can be related to the state-to-state rate coefficients by,
 	\begin{eqnarray}\label{Eq.inser}
 		k^{\mathrm{inser}}_{0j}(T) = \sum_{v'=0}^{4} \left(\sum_{j'=0,2,...}^{16} \frac{1}{2}\, k^{\mathrm{inser}}_{v'j',0,j} + \sum_{j'=1,3,...}^{17} \frac{3}{2}\, k^{\mathrm{inser}}_{v'j',0,j} \right),
 	\end{eqnarray} 
    where we summed over the final states of H$_2$, both vibrational ($v'=0,\dots,4$ were considered) and rotational. The sum over $j'$ is separated in ortho- and para-H$_2$ contributions and weighted with the appropriate nuclear spin statistics, because the high-level-of-theory calculations were performed assuming distinguishable nuclei and by applying the post-antisymmetrisation procedure\cite{WerfelliGhofran2015Ltrc}. We should mention that in the sum over $j'$ in Eq. (\ref{Eq.inser}), rate coefficients for $j' > 7$ were obtained by means of the quasi-classical trajectory (QCT) method. 
    
    	\begin{figure*}
 		\centering
 		\includegraphics[width=1.\linewidth]{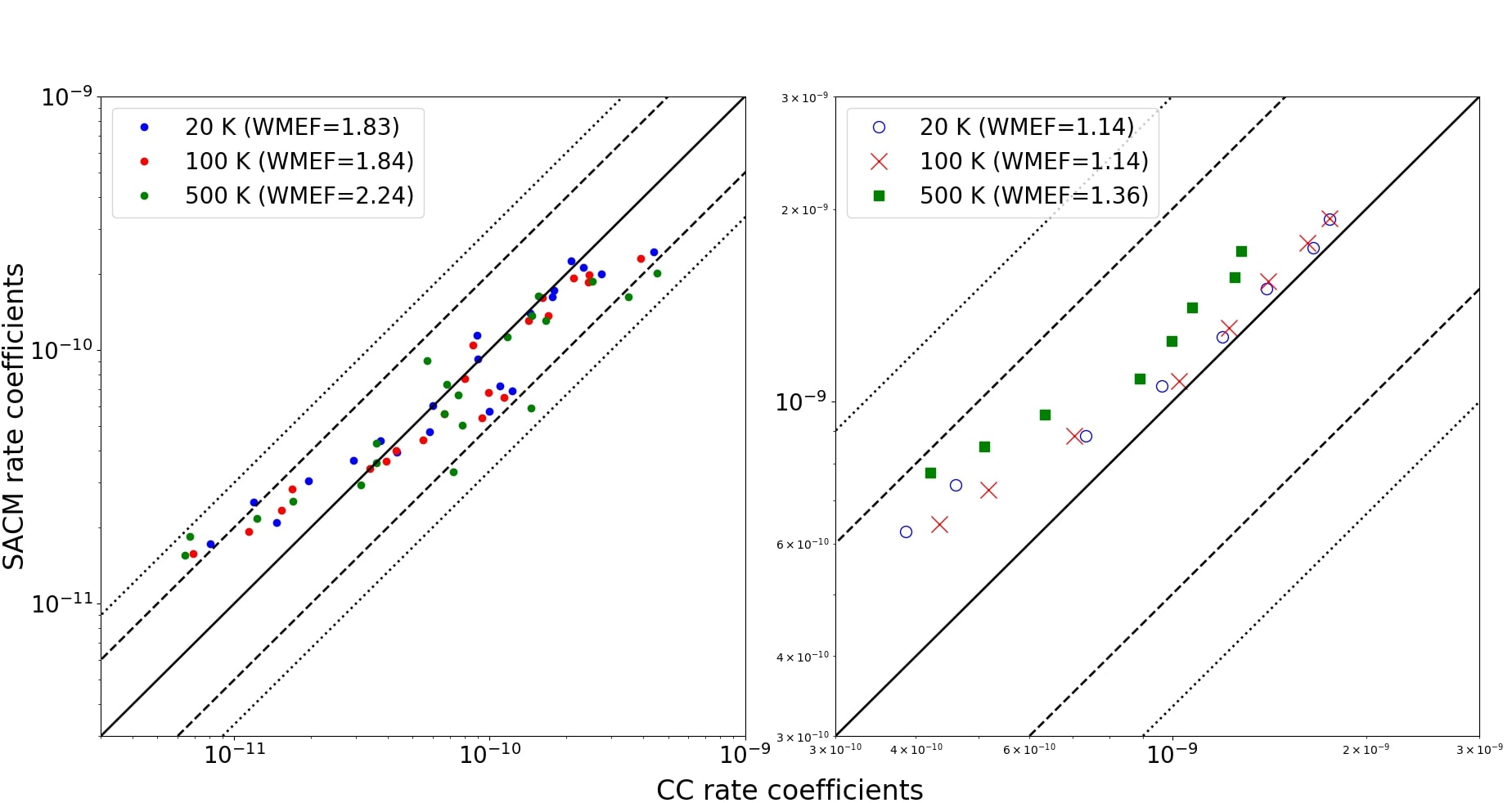}
 		\caption{Direct comparison at several temperatures of the SACM and the CC rate coefficients (in units of cm$^{3}$s$^{-1}$) for the rotationally inelastic collisions (left panel) and insertion reactions (right panel), involving CH$^+$ and H. The dashed (dotted) lines represent an error factor of 2 (3). The CC result were calculated by Werfelli \textit{et al.} (Ref. \onlinecite{WerfelliGhofran2015Ltrc}).}	
 		\label{fig:CH_Hp}
 	\end{figure*}
    
 	The SACM rate coefficients for the inelastic collision and insertion reaction were compared to the ones computed with the close-coupling method. The results for different temperatures are shown in Figure \ref{fig:CH_Hp}.
 	The agreement between the SACM rate coefficients and the CC results for the inelastic collisions is good at low temperatures. At higher temperatures, the accuracy as compared to state-to-state CC rates becomes significantly worse. An obvious reason for this discrepancy at higher temperatures could be the fact that the statistical approach is not supposed to work at high collision energies, which are mostly represented at high temperature. In addition, the authors of Ref. \onlinecite{WerfelliGhofran2015Ltrc} mention that they had difficulty converging the scattering cross sections and rate coefficients with respect to the $E_{\text{max}}$ parameter in the ABC code\cite{Skouteris2000} for the time-independent quantum scattering calculations. Consequently, these problems are reflected in the quality of the state-to-state CC rate coefficients for the inelastic collisions, especially at high temperature. 
 	The initial-state selected rate coefficients for the insertion reaction as computed with the close-coupling method do agree rather well with our statistical results, at least for the dominant initial rotational states of CH$^+$. 		
		
	\section{Conclusions and Outlook}\label{sec.4}	
		
	Computing state-to-state scattering cross sections and rate coefficients for (non-)reactive scattering events with highly accurate fully quantum mechanical approaches is no easy feat, especially when a long-lived intermediate collision complex is formed in the course of the collision. However, due to the statistical nature of these complex-mode processes, one can get away (typically) with much simpler statistical approaches to calculate these quantities, at least approximately so. In this article, we have showed that the statistical adiabatic channel model (SACM), relying upon the use of adiabatic potential curves as computed according to the improved way, provides an interesting and useful alternative to quantum approaches, not only for rotationally inelastic collisions as shown before (see Refs. \onlinecite{Loreau2018,Loreau_2018}), but also for combined rotational-vibrational transitions in inelastic collisions, as well as insertion reactions that are in competition with inelastic processes. Indeed, for the state-to-state rate coefficients of inelastic collisions and insertion reactions, we found an accuracy of better than a factor 2 for most transitions for the systems considered in this work, which is accurate enough for applications such as astrochemistry.
	The applicability of the current implementation of this SACM-inspired approach is however limited, since it requires the collision to happen exclusively via the formation and decay of a collision complex (see Eq. (\ref{Eq.2.1})). For some collisions, there might exist a competition between indirect (complex-mode) and direct transitions\cite{Gonzalez-Lezana20017,Miller1970,Quack&Troe3}. For such cases, the current implementation of the employed statistical quantum method is not expected to yield accurate results. For systems were such a competition is possible, it might be interesting to investigate the impact these direct transitions have on the statistical cross sections and rate coefficients in future research.
	Consequently, the benchmark systems considered in this work were all of the same type: a collision involving an atom and a diatomic molecule, characterized by a (mostly) completely attractive potential with a (deep) well. There are of course other potentially interesting complex-forming atom-diatom systems, which have been the subject of previous research (see Refs. \onlinecite{Rackham2001,Rackham2003,Aoiz2,Reyes,Reyes2,GONZALEZLEZANA2021138228,Yanan,Reyes3}), that could be studied with the statistical adiabatic channel model so as to extend the benchmarking even more. Examples include, but are not limited to: N ($^2D$) + H$_2$, {N ($^2D$) + D$_2$, O ($^1D$) + H$_2$, O ($^1D$) + HD, O ($^1D$) + D$_2$, C ($^1D$) + H$_2$, C ($^1D$) + D$_2$, C ($^1D$) + HD, and S ($^1D$) + HD.
	
	Even though only atom-diatom collisions were studied in this work, the applicability of the SACM-inspired approach is by no means limited to this type of low dimensional collisions. At time of writing the SACM has already successfully been applied to complex-mode inelastic collisions characterized by more than 3 internal degrees of freedom\cite{Loreau2018}. It would be desirable to do the same for reactive systems of (even) higher dimensionality.

	\section*{Conflicts of interest}	
	There are no conflicts of interest to declare.	
		
	\begin{acknowledgments}
	 We thank T. Stoecklin, T. Gonz\'{a}lez-Lezana, M. Lepers, P. Honvault, and A. Faure for sharing their data (PES or rate coefficients) used for comparison and for useful discussions.
	 J.L. acknowledges support from Internal Funds KU Leuven through grant STG-19-00313. 
	 The resources and services used in our computations were provided by the VSC (Flemish Supercomputer Center), funded by the Research Foundation–Flanders (FWO) and the Flemish Government. 
	 This project has received funding from the European Research Council (ERC) under the European Union's Horizon 2020 research and innovation programme (Grant agreement No. 811363). We acknowledge the Programme National Physique et Chimie du Milieu Interstellaire (PCMI) of CNRS/INSU with INC/INP co-funded by CEA and CNES. F.L. acknowledges financial support from the Institut Universitaire de France. 

	\end{acknowledgments}
	
    \section*{Data Availability}
	
	Data available on request from the authors.

		\bibliography{bibl}

\begin{thebibliography}{61}%
\makeatletter
\providecommand \@ifxundefined [1]{%
 \@ifx{#1\undefined}
}%
\providecommand \@ifnum [1]{%
 \ifnum #1\expandafter \@firstoftwo
 \else \expandafter \@secondoftwo
 \fi
}%
\providecommand \@ifx [1]{%
 \ifx #1\expandafter \@firstoftwo
 \else \expandafter \@secondoftwo
 \fi
}%
\providecommand \natexlab [1]{#1}%
\providecommand \enquote  [1]{``#1''}%
\providecommand \bibnamefont  [1]{#1}%
\providecommand \bibfnamefont [1]{#1}%
\providecommand \citenamefont [1]{#1}%
\providecommand \href@noop [0]{\@secondoftwo}%
\providecommand \href [0]{\begingroup \@sanitize@url \@href}%
\providecommand \@href[1]{\@@startlink{#1}\@@href}%
\providecommand \@@href[1]{\endgroup#1\@@endlink}%
\providecommand \@sanitize@url [0]{\catcode `\\12\catcode `\$12\catcode
  `\&12\catcode `\#12\catcode `\^12\catcode `\_12\catcode `\%12\relax}%
\providecommand \@@startlink[1]{}%
\providecommand \@@endlink[0]{}%
\providecommand \url  [0]{\begingroup\@sanitize@url \@url }%
\providecommand \@url [1]{\endgroup\@href {#1}{\urlprefix }}%
\providecommand \urlprefix  [0]{URL }%
\providecommand \Eprint [0]{\href }%
\providecommand \doibase [0]{http://dx.doi.org/}%
\providecommand \selectlanguage [0]{\@gobble}%
\providecommand \bibinfo  [0]{\@secondoftwo}%
\providecommand \bibfield  [0]{\@secondoftwo}%
\providecommand \translation [1]{[#1]}%
\providecommand \BibitemOpen [0]{}%
\providecommand \bibitemStop [0]{}%
\providecommand \bibitemNoStop [0]{.\EOS\space}%
\providecommand \EOS [0]{\spacefactor3000\relax}%
\providecommand \BibitemShut  [1]{\csname bibitem#1\endcsname}%
\let\auto@bib@innerbib\@empty
\bibitem [{\citenamefont {Faure}, \citenamefont {Lique},\ and\ \citenamefont
  {Loreau}(2020)}]{Faure2020}%
  \BibitemOpen
  \bibfield  {author} {\bibinfo {author} {\bibfnamefont {A.}~\bibnamefont
  {Faure}}, \bibinfo {author} {\bibfnamefont {F.}~\bibnamefont {Lique}}, \ and\
  \bibinfo {author} {\bibfnamefont {J.}~\bibnamefont {Loreau}},\ }\href
  {\doibase 10.1093/mnras/staa242} {\bibfield  {journal} {\bibinfo  {journal}
  {Monthly Notices of the Royal Astronomical Society}\ }\textbf {\bibinfo
  {volume} {493}},\ \bibinfo {pages} {776–782} (\bibinfo {year}
  {2020})}\BibitemShut {NoStop}%
\bibitem [{\citenamefont {Zhang}\ and\ \citenamefont {Zhang}(1994)}]{Zhang}%
  \BibitemOpen
  \bibfield  {author} {\bibinfo {author} {\bibfnamefont {D.~H.}\ \bibnamefont
  {Zhang}}\ and\ \bibinfo {author} {\bibfnamefont {J.~Z.~H.}\ \bibnamefont
  {Zhang}},\ }\href {\doibase 10.1063/1.467808} {\bibfield  {journal} {\bibinfo
   {journal} {The Journal of Chemical Physics}\ }\textbf {\bibinfo {volume}
  {101}},\ \bibinfo {pages} {1146--1156} (\bibinfo {year} {1994})}\BibitemShut
  {NoStop}%
\bibitem [{\citenamefont {Althorpe}\ and\ \citenamefont
  {Clary}(2003)}]{Althorpe}%
  \BibitemOpen
  \bibfield  {author} {\bibinfo {author} {\bibfnamefont {S.~C.}\ \bibnamefont
  {Althorpe}}\ and\ \bibinfo {author} {\bibfnamefont {D.~C.}\ \bibnamefont
  {Clary}},\ }\href {\doibase 10.1146/annurev.physchem.54.011002.103750}
  {\bibfield  {journal} {\bibinfo  {journal} {Annual Review of Physical
  Chemistry}\ }\textbf {\bibinfo {volume} {54}},\ \bibinfo {pages} {493--529}
  (\bibinfo {year} {2003})}\BibitemShut {NoStop}%
\bibitem [{\citenamefont {Honvault}\ and\ \citenamefont
  {Launay}(2004)}]{Honvault_rev}%
  \BibitemOpen
  \bibfield  {author} {\bibinfo {author} {\bibfnamefont {P.}~\bibnamefont
  {Honvault}}\ and\ \bibinfo {author} {\bibfnamefont {J.-M.}\ \bibnamefont
  {Launay}},\ }in\ \href@noop {} {\emph {\bibinfo {booktitle} {Theory of
  Chemical Reaction Dynamics}}},\ \bibinfo {editor} {edited by\ \bibinfo
  {editor} {\bibfnamefont {A.}~\bibnamefont {Lagana}}\ and\ \bibinfo {editor}
  {\bibfnamefont {G.}~\bibnamefont {Lendvay}}}\ (\bibinfo  {publisher}
  {Springer Netherlands},\ \bibinfo {address} {Dordrecht},\ \bibinfo {year}
  {2004})\ pp.\ \bibinfo {pages} {187--215}\BibitemShut {NoStop}%
\bibitem [{\citenamefont {González-Lezana}(2007)}]{Gonzalez-Lezana20017}%
  \BibitemOpen
  \bibfield  {author} {\bibinfo {author} {\bibfnamefont {T.}~\bibnamefont
  {González-Lezana}},\ }\href {\doibase 10.1080/03081070600933476} {\bibfield
  {journal} {\bibinfo  {journal} {International Reviews in Physical Chemistry}\
  }\textbf {\bibinfo {volume} {26}},\ \bibinfo {pages} {29--91} (\bibinfo
  {year} {2007})}\BibitemShut {NoStop}%
\bibitem [{\citenamefont {Loreau}, \citenamefont {Lique},\ and\ \citenamefont
  {Faure}(2018)}]{Loreau_2018}%
  \BibitemOpen
  \bibfield  {author} {\bibinfo {author} {\bibfnamefont {J.}~\bibnamefont
  {Loreau}}, \bibinfo {author} {\bibfnamefont {F.}~\bibnamefont {Lique}}, \
  and\ \bibinfo {author} {\bibfnamefont {A.}~\bibnamefont {Faure}},\ }\href
  {\doibase 10.3847/2041-8213/aaa5fe} {\bibfield  {journal} {\bibinfo
  {journal} {The Astrophysical Journal}\ }\textbf {\bibinfo {volume} {853}},\
  \bibinfo {pages} {L5} (\bibinfo {year} {2018})}\BibitemShut {NoStop}%
\bibitem [{\citenamefont {Light}(1967)}]{DF9674400014}%
  \BibitemOpen
  \bibfield  {author} {\bibinfo {author} {\bibfnamefont {J.~C.}\ \bibnamefont
  {Light}},\ }\href {\doibase 10.1039/DF9674400014} {\bibfield  {journal}
  {\bibinfo  {journal} {Discuss. Faraday Soc.}\ }\textbf {\bibinfo {volume}
  {44}},\ \bibinfo {pages} {14--29} (\bibinfo {year} {1967})}\BibitemShut
  {NoStop}%
\bibitem [{\citenamefont {Bernstein}\ \emph {et~al.}(1963)\citenamefont
  {Bernstein}, \citenamefont {Dalgarno}, \citenamefont {Massey},\ and\
  \citenamefont {Percival}}]{Bernstein1963}%
  \BibitemOpen
  \bibfield  {author} {\bibinfo {author} {\bibfnamefont {R.~B.}\ \bibnamefont
  {Bernstein}}, \bibinfo {author} {\bibfnamefont {A.}~\bibnamefont {Dalgarno}},
  \bibinfo {author} {\bibfnamefont {H.~S.~W.}\ \bibnamefont {Massey}}, \ and\
  \bibinfo {author} {\bibfnamefont {I.~C.}\ \bibnamefont {Percival}},\ }\href
  {\doibase 10.1098/rspa.1963.0142} {\bibfield  {journal} {\bibinfo  {journal}
  {Proceedings of the Royal Society of London. Series A. Mathematical and
  Physical Sciences}\ }\textbf {\bibinfo {volume} {274}},\ \bibinfo {pages}
  {427--442} (\bibinfo {year} {1963})}\BibitemShut {NoStop}%
\bibitem [{\citenamefont {Pechukas}, \citenamefont {Light},\ and\ \citenamefont
  {Rankin}(1966)}]{PechukasLight1964}%
  \BibitemOpen
  \bibfield  {author} {\bibinfo {author} {\bibfnamefont {P.}~\bibnamefont
  {Pechukas}}, \bibinfo {author} {\bibfnamefont {J.~C.}\ \bibnamefont {Light}},
  \ and\ \bibinfo {author} {\bibfnamefont {C.}~\bibnamefont {Rankin}},\ }\href
  {\doibase 10.1063/1.1726760} {\bibfield  {journal} {\bibinfo  {journal} {The
  Journal of Chemical Physics}\ }\textbf {\bibinfo {volume} {44}},\ \bibinfo
  {pages} {794--805} (\bibinfo {year} {1966})}\BibitemShut {NoStop}%
\bibitem [{\citenamefont {Pechukas}\ and\ \citenamefont
  {Light}(1965)}]{PechukasLight1965}%
  \BibitemOpen
  \bibfield  {author} {\bibinfo {author} {\bibfnamefont {P.}~\bibnamefont
  {Pechukas}}\ and\ \bibinfo {author} {\bibfnamefont {J.~C.}\ \bibnamefont
  {Light}},\ }\href {\doibase 10.1063/1.1696411} {\bibfield  {journal}
  {\bibinfo  {journal} {The Journal of Chemical Physics}\ }\textbf {\bibinfo
  {volume} {42}},\ \bibinfo {pages} {3281--3291} (\bibinfo {year}
  {1965})}\BibitemShut {NoStop}%
\bibitem [{\citenamefont {Miller}(1970)}]{Miller1970}%
  \BibitemOpen
  \bibfield  {author} {\bibinfo {author} {\bibfnamefont {W.~H.}\ \bibnamefont
  {Miller}},\ }\href {\doibase 10.1063/1.1673020} {\bibfield  {journal}
  {\bibinfo  {journal} {The Journal of Chemical Physics}\ }\textbf {\bibinfo
  {volume} {52}},\ \bibinfo {pages} {543--551} (\bibinfo {year}
  {1970})}\BibitemShut {NoStop}%
\bibitem [{\citenamefont {Larr\'{e}garay}, \citenamefont {Bonnet},\ and\
  \citenamefont {Rayez}(2007)}]{Larregaray}%
  \BibitemOpen
  \bibfield  {author} {\bibinfo {author} {\bibfnamefont {P.}~\bibnamefont
  {Larr\'{e}garay}}, \bibinfo {author} {\bibfnamefont {L.}~\bibnamefont
  {Bonnet}}, \ and\ \bibinfo {author} {\bibfnamefont {J.-C.}\ \bibnamefont
  {Rayez}},\ }\href {\doibase 10.1063/1.2768959} {\bibfield  {journal}
  {\bibinfo  {journal} {The Journal of Chemical Physics}\ }\textbf {\bibinfo
  {volume} {127}},\ \bibinfo {pages} {084308} (\bibinfo {year}
  {2007})}\BibitemShut {NoStop}%
\bibitem [{\citenamefont {Larr\'{e}garay}, \citenamefont {Bonnet},\ and\
  \citenamefont {Rayez}(2006)}]{Bonnet}%
  \BibitemOpen
  \bibfield  {author} {\bibinfo {author} {\bibfnamefont {P.}~\bibnamefont
  {Larr\'{e}garay}}, \bibinfo {author} {\bibfnamefont {L.}~\bibnamefont
  {Bonnet}}, \ and\ \bibinfo {author} {\bibfnamefont {J.-C.}\ \bibnamefont
  {Rayez}},\ }\href {\doibase 10.1021/jp053822x} {\bibfield  {journal}
  {\bibinfo  {journal} {The Journal of Physical Chemistry A}\ }\textbf
  {\bibinfo {volume} {110}},\ \bibinfo {pages} {1552--1560} (\bibinfo {year}
  {2006})}\BibitemShut {NoStop}%
\bibitem [{\citenamefont {Bonnet}, \citenamefont {Larr\'{e}garay},\ and\
  \citenamefont {Rayez}(2007)}]{B700906B}%
  \BibitemOpen
  \bibfield  {author} {\bibinfo {author} {\bibfnamefont {L.}~\bibnamefont
  {Bonnet}}, \bibinfo {author} {\bibfnamefont {P.}~\bibnamefont
  {Larr\'{e}garay}}, \ and\ \bibinfo {author} {\bibfnamefont {J.-C.}\
  \bibnamefont {Rayez}},\ }\href {\doibase 10.1039/B700906B} {\bibfield
  {journal} {\bibinfo  {journal} {Phys. Chem. Chem. Phys.}\ }\textbf {\bibinfo
  {volume} {9}},\ \bibinfo {pages} {3228--3240} (\bibinfo {year}
  {2007})}\BibitemShut {NoStop}%
\bibitem [{\citenamefont {Rackham}, \citenamefont {Huarte-Larranaga},\ and\
  \citenamefont {Manolopoulos}(2001)}]{Rackham2001}%
  \BibitemOpen
  \bibfield  {author} {\bibinfo {author} {\bibfnamefont {E.~J.}\ \bibnamefont
  {Rackham}}, \bibinfo {author} {\bibfnamefont {F.}~\bibnamefont
  {Huarte-Larranaga}}, \ and\ \bibinfo {author} {\bibfnamefont {D.~E.}\
  \bibnamefont {Manolopoulos}},\ }\href {\doibase
  https://doi.org/10.1016/S0009-2614(01)00707-2} {\bibfield  {journal}
  {\bibinfo  {journal} {Chemical Physics Letters}\ }\textbf {\bibinfo {volume}
  {343}},\ \bibinfo {pages} {356 -- 364} (\bibinfo {year} {2001})}\BibitemShut
  {NoStop}%
\bibitem [{\citenamefont {Rackham}, \citenamefont {Gonzalez-Lezana},\ and\
  \citenamefont {Manolopoulos}(2003)}]{Rackham2003}%
  \BibitemOpen
  \bibfield  {author} {\bibinfo {author} {\bibfnamefont {E.~J.}\ \bibnamefont
  {Rackham}}, \bibinfo {author} {\bibfnamefont {T.}~\bibnamefont
  {Gonzalez-Lezana}}, \ and\ \bibinfo {author} {\bibfnamefont {D.~E.}\
  \bibnamefont {Manolopoulos}},\ }\href {\doibase 10.1063/1.1628218} {\bibfield
   {journal} {\bibinfo  {journal} {The Journal of Chemical Physics}\ }\textbf
  {\bibinfo {volume} {119}},\ \bibinfo {pages} {12895--12907} (\bibinfo {year}
  {2003})}\BibitemShut {NoStop}%
\bibitem [{\citenamefont {Clary}\ and\ \citenamefont {Henshaw}(1987)}]{Clary}%
  \BibitemOpen
  \bibfield  {author} {\bibinfo {author} {\bibfnamefont {D.~C.}\ \bibnamefont
  {Clary}}\ and\ \bibinfo {author} {\bibfnamefont {J.~P.}\ \bibnamefont
  {Henshaw}},\ }\href {\doibase 10.1039/DC9878400333} {\bibfield  {journal}
  {\bibinfo  {journal} {Faraday Discuss. Chem. Soc.}\ }\textbf {\bibinfo
  {volume} {84}},\ \bibinfo {pages} {333--349} (\bibinfo {year}
  {1987})}\BibitemShut {NoStop}%
\bibitem [{\citenamefont {Dagdigian}\ and\ \citenamefont
  {Alexander}(2018)}]{Dagdigian}%
  \BibitemOpen
  \bibfield  {author} {\bibinfo {author} {\bibfnamefont {P.~J.}\ \bibnamefont
  {Dagdigian}}\ and\ \bibinfo {author} {\bibfnamefont {M.~H.}\ \bibnamefont
  {Alexander}}\ }(\bibinfo  {publisher} {John Wiley \& Sons, Ltd},\ \bibinfo
  {year} {2018})\ Chap.~\bibinfo {chapter} {1}, pp.\ \bibinfo {pages}
  {1--43}\BibitemShut {NoStop}%
\bibitem [{\citenamefont {Aoiz}\ \emph {et~al.}(2007)\citenamefont {Aoiz},
  \citenamefont {Sáez~Rábanos}, \citenamefont {González-Lezana},\ and\
  \citenamefont {Manolopoulos}}]{Aoiz}%
  \BibitemOpen
  \bibfield  {author} {\bibinfo {author} {\bibfnamefont {F.~J.}\ \bibnamefont
  {Aoiz}}, \bibinfo {author} {\bibfnamefont {V.}~\bibnamefont
  {Sáez~Rábanos}}, \bibinfo {author} {\bibfnamefont {T.}~\bibnamefont
  {González-Lezana}}, \ and\ \bibinfo {author} {\bibfnamefont {D.~E.}\
  \bibnamefont {Manolopoulos}},\ }\href {\doibase 10.1063/1.2723067} {\bibfield
   {journal} {\bibinfo  {journal} {The Journal of Chemical Physics}\ }\textbf
  {\bibinfo {volume} {126}},\ \bibinfo {pages} {161101} (\bibinfo {year}
  {2007})}\BibitemShut {NoStop}%
\bibitem [{\citenamefont {Aoiz}, \citenamefont {González-Lezana},\ and\
  \citenamefont {Sáez~Rábanos}(2008)}]{Aoiz2}%
  \BibitemOpen
  \bibfield  {author} {\bibinfo {author} {\bibfnamefont {F.~J.}\ \bibnamefont
  {Aoiz}}, \bibinfo {author} {\bibfnamefont {T.}~\bibnamefont
  {González-Lezana}}, \ and\ \bibinfo {author} {\bibfnamefont
  {V.}~\bibnamefont {Sáez~Rábanos}},\ }\href {\doibase 10.1063/1.2969812}
  {\bibfield  {journal} {\bibinfo  {journal} {The Journal of Chemical Physics}\
  }\textbf {\bibinfo {volume} {129}},\ \bibinfo {pages} {094305} (\bibinfo
  {year} {2008})}\BibitemShut {NoStop}%
\bibitem [{\citenamefont {Jambrina}\ \emph {et~al.}(2009)\citenamefont
  {Jambrina}, \citenamefont {Aoiz}, \citenamefont {Eyles}, \citenamefont
  {Herrero},\ and\ \citenamefont {Sáez~Rábanos}}]{Jambrina}%
  \BibitemOpen
  \bibfield  {author} {\bibinfo {author} {\bibfnamefont {P.~G.}\ \bibnamefont
  {Jambrina}}, \bibinfo {author} {\bibfnamefont {F.~J.}\ \bibnamefont {Aoiz}},
  \bibinfo {author} {\bibfnamefont {C.~J.}\ \bibnamefont {Eyles}}, \bibinfo
  {author} {\bibfnamefont {V.~J.}\ \bibnamefont {Herrero}}, \ and\ \bibinfo
  {author} {\bibfnamefont {V.}~\bibnamefont {Sáez~Rábanos}},\ }\href
  {\doibase 10.1063/1.3129343} {\bibfield  {journal} {\bibinfo  {journal} {The
  Journal of Chemical Physics}\ }\textbf {\bibinfo {volume} {130}},\ \bibinfo
  {pages} {184303} (\bibinfo {year} {2009})}\BibitemShut {NoStop}%
\bibitem [{\citenamefont {Quack}\ and\ \citenamefont
  {Troe}(1974)}]{Quack&Troe2}%
  \BibitemOpen
  \bibfield  {author} {\bibinfo {author} {\bibfnamefont {M.}~\bibnamefont
  {Quack}}\ and\ \bibinfo {author} {\bibfnamefont {J.}~\bibnamefont {Troe}},\
  }\href {\doibase 10.1002/bbpc.19740780306} {\bibfield  {journal} {\bibinfo
  {journal} {Berichte der Bunsengesellschaft f\"{u}r physikalische Chemie}\
  }\textbf {\bibinfo {volume} {78}},\ \bibinfo {pages} {240--252} (\bibinfo
  {year} {1974})}\BibitemShut {NoStop}%
\bibitem [{\citenamefont {Quack}\ and\ \citenamefont
  {Troe}(1975)}]{Quack&Troe3}%
  \BibitemOpen
  \bibfield  {author} {\bibinfo {author} {\bibfnamefont {M.}~\bibnamefont
  {Quack}}\ and\ \bibinfo {author} {\bibfnamefont {J.}~\bibnamefont {Troe}},\
  }\href {\doibase 10.1002/bbpc.19750790211} {\bibfield  {journal} {\bibinfo
  {journal} {Berichte der Bunsengesellschaft f\"{u}r physikalische Chemie}\
  }\textbf {\bibinfo {volume} {79}},\ \bibinfo {pages} {170--183} (\bibinfo
  {year} {1975})}\BibitemShut {NoStop}%
\bibitem [{\citenamefont {Quack}\ and\ \citenamefont
  {Troe}(1976)}]{Quack&Troe_inform}%
  \BibitemOpen
  \bibfield  {author} {\bibinfo {author} {\bibfnamefont {M.}~\bibnamefont
  {Quack}}\ and\ \bibinfo {author} {\bibfnamefont {J.}~\bibnamefont {Troe}},\
  }\href {\doibase https://doi.org/10.1002/bbpc.19760801112} {\bibfield
  {journal} {\bibinfo  {journal} {Berichte der Bunsengesellschaft für
  physikalische Chemie}\ }\textbf {\bibinfo {volume} {80}},\ \bibinfo {pages}
  {1140--1149} (\bibinfo {year} {1976})}\BibitemShut {NoStop}%
\bibitem [{\citenamefont {Clary}(1987)}]{Clary_ad}%
  \BibitemOpen
  \bibfield  {author} {\bibinfo {author} {\bibfnamefont {D.~C.}\ \bibnamefont
  {Clary}},\ }\href {\doibase 10.1039/F29878300139} {\bibfield  {journal}
  {\bibinfo  {journal} {J. Chem. Soc.{,} Faraday Trans. 2}\ }\textbf {\bibinfo
  {volume} {83}},\ \bibinfo {pages} {139--148} (\bibinfo {year}
  {1987})}\BibitemShut {NoStop}%
\bibitem [{\citenamefont {Smith}\ and\ \citenamefont {Troe}(1992)}]{Smith}%
  \BibitemOpen
  \bibfield  {author} {\bibinfo {author} {\bibfnamefont {S.~C.}\ \bibnamefont
  {Smith}}\ and\ \bibinfo {author} {\bibfnamefont {J.}~\bibnamefont {Troe}},\
  }\href {\doibase 10.1063/1.463804} {\bibfield  {journal} {\bibinfo  {journal}
  {The Journal of Chemical Physics}\ }\textbf {\bibinfo {volume} {97}},\
  \bibinfo {pages} {5451--5464} (\bibinfo {year} {1992})}\BibitemShut {NoStop}%
\bibitem [{\citenamefont {Loreau}, \citenamefont {Faure},\ and\ \citenamefont
  {Lique}(2018)}]{Loreau2018}%
  \BibitemOpen
  \bibfield  {author} {\bibinfo {author} {\bibfnamefont {J.}~\bibnamefont
  {Loreau}}, \bibinfo {author} {\bibfnamefont {A.}~\bibnamefont {Faure}}, \
  and\ \bibinfo {author} {\bibfnamefont {F.}~\bibnamefont {Lique}},\ }\href
  {\doibase 10.1063/1.5036819} {\bibfield  {journal} {\bibinfo  {journal} {The
  Journal of Chemical Physics}\ }\textbf {\bibinfo {volume} {148}},\ \bibinfo
  {pages} {244308} (\bibinfo {year} {2018})}\BibitemShut {NoStop}%
\bibitem [{\citenamefont {Quack}(1977)}]{QuackSymm}%
  \BibitemOpen
  \bibfield  {author} {\bibinfo {author} {\bibfnamefont {M.}~\bibnamefont
  {Quack}},\ }\href {\doibase 10.1080/00268977700101861} {\bibfield  {journal}
  {\bibinfo  {journal} {Molecular Physics}\ }\textbf {\bibinfo {volume} {34}},\
  \bibinfo {pages} {477--504} (\bibinfo {year} {1977})}\BibitemShut {NoStop}%
\bibitem [{\citenamefont {Lique}, \citenamefont {Honvault},\ and\ \citenamefont
  {Faure}(2012)}]{Lique2012}%
  \BibitemOpen
  \bibfield  {author} {\bibinfo {author} {\bibfnamefont {F.}~\bibnamefont
  {Lique}}, \bibinfo {author} {\bibfnamefont {P.}~\bibnamefont {Honvault}}, \
  and\ \bibinfo {author} {\bibfnamefont {A.}~\bibnamefont {Faure}},\ }\href
  {\doibase 10.1063/1.4758791} {\bibfield  {journal} {\bibinfo  {journal} {The
  Journal of Chemical Physics}\ }\textbf {\bibinfo {volume} {137}},\ \bibinfo
  {pages} {154303} (\bibinfo {year} {2012})}\BibitemShut {NoStop}%
\bibitem [{\citenamefont {González-Lezana}, \citenamefont {Hily-Blant},\ and\
  \citenamefont {Faure}(2021)}]{H3pSQM}%
  \BibitemOpen
  \bibfield  {author} {\bibinfo {author} {\bibfnamefont {T.}~\bibnamefont
  {González-Lezana}}, \bibinfo {author} {\bibfnamefont {P.}~\bibnamefont
  {Hily-Blant}}, \ and\ \bibinfo {author} {\bibfnamefont {A.}~\bibnamefont
  {Faure}},\ }\href {\doibase 10.1063/5.0039629} {\bibfield  {journal}
  {\bibinfo  {journal} {The Journal of Chemical Physics}\ }\textbf {\bibinfo
  {volume} {154}},\ \bibinfo {pages} {054310} (\bibinfo {year}
  {2021})}\BibitemShut {NoStop}%
\bibitem [{\citenamefont {Miller}(1969)}]{Miller1969}%
  \BibitemOpen
  \bibfield  {author} {\bibinfo {author} {\bibfnamefont {W.~H.}\ \bibnamefont
  {Miller}},\ }\href {\doibase 10.1063/1.1670812} {\bibfield  {journal}
  {\bibinfo  {journal} {The Journal of Chemical Physics}\ }\textbf {\bibinfo
  {volume} {50}},\ \bibinfo {pages} {407--418} (\bibinfo {year}
  {1969})}\BibitemShut {NoStop}%
\bibitem [{\citenamefont {Grozdanov}\ and\ \citenamefont
  {McCarroll}(2012)}]{Grozdanov}%
  \BibitemOpen
  \bibfield  {author} {\bibinfo {author} {\bibfnamefont {T.~P.}\ \bibnamefont
  {Grozdanov}}\ and\ \bibinfo {author} {\bibfnamefont {R.}~\bibnamefont
  {McCarroll}},\ }\href {\doibase 10.1021/jp210992g} {\bibfield  {journal}
  {\bibinfo  {journal} {The Journal of Physical Chemistry A}\ }\textbf
  {\bibinfo {volume} {116}},\ \bibinfo {pages} {4569--4577} (\bibinfo {year}
  {2012})}\BibitemShut {NoStop}%
\bibitem [{\citenamefont {Park}\ and\ \citenamefont
  {Light}(2007)}]{Park&Light}%
  \BibitemOpen
  \bibfield  {author} {\bibinfo {author} {\bibfnamefont {K.}~\bibnamefont
  {Park}}\ and\ \bibinfo {author} {\bibfnamefont {J.~C.}\ \bibnamefont
  {Light}},\ }\href {\doibase 10.1063/1.2805394} {\bibfield  {journal}
  {\bibinfo  {journal} {The Journal of Chemical Physics}\ }\textbf {\bibinfo
  {volume} {127}},\ \bibinfo {pages} {224101} (\bibinfo {year}
  {2007})}\BibitemShut {NoStop}%
\bibitem [{\citenamefont {Honvault}\ \emph {et~al.}(2011)\citenamefont
  {Honvault}, \citenamefont {Jorfi}, \citenamefont {González-Lezana},
  \citenamefont {Faure},\ and\ \citenamefont {Pagani}}]{HonvaultP2011OHcb}%
  \BibitemOpen
  \bibfield  {author} {\bibinfo {author} {\bibfnamefont {P.}~\bibnamefont
  {Honvault}}, \bibinfo {author} {\bibfnamefont {M.}~\bibnamefont {Jorfi}},
  \bibinfo {author} {\bibfnamefont {T.}~\bibnamefont {González-Lezana}},
  \bibinfo {author} {\bibfnamefont {A.}~\bibnamefont {Faure}}, \ and\ \bibinfo
  {author} {\bibfnamefont {L.}~\bibnamefont {Pagani}},\ }\href@noop {}
  {\bibfield  {journal} {\bibinfo  {journal} {Physical review letters}\
  }\textbf {\bibinfo {volume} {107}},\ \bibinfo {pages} {023201--023201}
  (\bibinfo {year} {2011})}\BibitemShut {NoStop}%
\bibitem [{\citenamefont {Flower}(2007)}]{Flower2007}%
  \BibitemOpen
  \bibfield  {author} {\bibinfo {author} {\bibfnamefont {D.}~\bibnamefont
  {Flower}}\ }(\bibinfo  {publisher} {Cambridge University Press},\ \bibinfo
  {year} {2007})\ p.\ \bibinfo {pages} {118–138},\ \bibinfo {edition} {2nd}\
  ed.\BibitemShut {Stop}%
\bibitem [{\citenamefont {Hutson}\ and\ \citenamefont
  {Le~Sueur}(2019)}]{HutsonJeremyM2019mApf}%
  \BibitemOpen
  \bibfield  {author} {\bibinfo {author} {\bibfnamefont {J.~M.}\ \bibnamefont
  {Hutson}}\ and\ \bibinfo {author} {\bibfnamefont {C.~R.}\ \bibnamefont
  {Le~Sueur}},\ }\href@noop {} {\bibfield  {journal} {\bibinfo  {journal}
  {Computer Physics Communications}\ }\textbf {\bibinfo {volume} {241}},\
  \bibinfo {pages} {9--18} (\bibinfo {year} {2019})}\BibitemShut {NoStop}%
\bibitem [{\citenamefont {Dashevskaya}\ \emph {et~al.}(2003)\citenamefont
  {Dashevskaya}, \citenamefont {Maergoiz}, \citenamefont {Troe}, \citenamefont
  {Litvin},\ and\ \citenamefont {Nikitin}}]{Dashevskaya}%
  \BibitemOpen
  \bibfield  {author} {\bibinfo {author} {\bibfnamefont {E.~I.}\ \bibnamefont
  {Dashevskaya}}, \bibinfo {author} {\bibfnamefont {A.~I.}\ \bibnamefont
  {Maergoiz}}, \bibinfo {author} {\bibfnamefont {J.}~\bibnamefont {Troe}},
  \bibinfo {author} {\bibfnamefont {I.}~\bibnamefont {Litvin}}, \ and\ \bibinfo
  {author} {\bibfnamefont {E.~E.}\ \bibnamefont {Nikitin}},\ }\href {\doibase
  10.1063/1.1562159} {\bibfield  {journal} {\bibinfo  {journal} {The Journal of
  Chemical Physics}\ }\textbf {\bibinfo {volume} {118}},\ \bibinfo {pages}
  {7313--7320} (\bibinfo {year} {2003})}\BibitemShut {NoStop}%
\bibitem [{\citenamefont {Velilla}\ \emph {et~al.}(2008)\citenamefont
  {Velilla}, \citenamefont {Lepetit}, \citenamefont {Aguado}, \citenamefont
  {Beswick},\ and\ \citenamefont {Paniagua}}]{Velilla}%
  \BibitemOpen
  \bibfield  {author} {\bibinfo {author} {\bibfnamefont {L.}~\bibnamefont
  {Velilla}}, \bibinfo {author} {\bibfnamefont {B.}~\bibnamefont {Lepetit}},
  \bibinfo {author} {\bibfnamefont {A.}~\bibnamefont {Aguado}}, \bibinfo
  {author} {\bibfnamefont {J.~A.}\ \bibnamefont {Beswick}}, \ and\ \bibinfo
  {author} {\bibfnamefont {M.}~\bibnamefont {Paniagua}},\ }\href {\doibase
  10.1063/1.2973629} {\bibfield  {journal} {\bibinfo  {journal} {The Journal of
  Chemical Physics}\ }\textbf {\bibinfo {volume} {129}},\ \bibinfo {pages}
  {084307} (\bibinfo {year} {2008})}\BibitemShut {NoStop}%
\bibitem [{\citenamefont {Gerlich}(1990)}]{Gerlich}%
  \BibitemOpen
  \bibfield  {author} {\bibinfo {author} {\bibfnamefont {D.}~\bibnamefont
  {Gerlich}},\ }\href {\doibase 10.1063/1.457980} {\bibfield  {journal}
  {\bibinfo  {journal} {The Journal of Chemical Physics}\ }\textbf {\bibinfo
  {volume} {92}},\ \bibinfo {pages} {2377--2388} (\bibinfo {year}
  {1990})}\BibitemShut {NoStop}%
\bibitem [{\citenamefont {Gonz\'{a}lez-Lezana}\ \emph
  {et~al.}(2006)\citenamefont {Gonz\'{a}lez-Lezana}, \citenamefont {Roncero},
  \citenamefont {Honvault}, \citenamefont {Launay}, \citenamefont {Bulut},
  \citenamefont {Javier~Aoiz},\ and\ \citenamefont {Banares}}]{Lezana}%
  \BibitemOpen
  \bibfield  {author} {\bibinfo {author} {\bibfnamefont {T.}~\bibnamefont
  {Gonz\'{a}lez-Lezana}}, \bibinfo {author} {\bibfnamefont {O.}~\bibnamefont
  {Roncero}}, \bibinfo {author} {\bibfnamefont {P.}~\bibnamefont {Honvault}},
  \bibinfo {author} {\bibfnamefont {J.-M.}\ \bibnamefont {Launay}}, \bibinfo
  {author} {\bibfnamefont {N.}~\bibnamefont {Bulut}}, \bibinfo {author}
  {\bibfnamefont {F.}~\bibnamefont {Javier~Aoiz}}, \ and\ \bibinfo {author}
  {\bibfnamefont {L.}~\bibnamefont {Banares}},\ }\href {\doibase
  10.1063/1.2336224} {\bibfield  {journal} {\bibinfo  {journal} {The Journal of
  Chemical Physics}\ }\textbf {\bibinfo {volume} {125}},\ \bibinfo {pages}
  {094314} (\bibinfo {year} {2006})}\BibitemShut {NoStop}%
\bibitem [{\citenamefont {González-Lezana}\ and\ \citenamefont
  {Honvault}(2017)}]{Gonzalez-LezanaT2017}%
  \BibitemOpen
  \bibfield  {author} {\bibinfo {author} {\bibfnamefont {T.}~\bibnamefont
  {González-Lezana}}\ and\ \bibinfo {author} {\bibfnamefont {P.}~\bibnamefont
  {Honvault}},\ }\href@noop {} {\bibfield  {journal} {\bibinfo  {journal}
  {Monthly Notices of the Royal Astronomical Society}\ }\textbf {\bibinfo
  {volume} {467}},\ \bibinfo {pages} {1294--1299} (\bibinfo {year}
  {2017})}\BibitemShut {NoStop}%
\bibitem [{\citenamefont {Honvault}\ \emph {et~al.}(2012)\citenamefont
  {Honvault}, \citenamefont {Jorfi}, \citenamefont {Gonz\'alez-Lezana},
  \citenamefont {Faure},\ and\ \citenamefont {Pagani}}]{Erratum_Honv}%
  \BibitemOpen
  \bibfield  {author} {\bibinfo {author} {\bibfnamefont {P.}~\bibnamefont
  {Honvault}}, \bibinfo {author} {\bibfnamefont {M.}~\bibnamefont {Jorfi}},
  \bibinfo {author} {\bibfnamefont {T.}~\bibnamefont {Gonz\'alez-Lezana}},
  \bibinfo {author} {\bibfnamefont {A.}~\bibnamefont {Faure}}, \ and\ \bibinfo
  {author} {\bibfnamefont {L.}~\bibnamefont {Pagani}},\ }\href {\doibase
  10.1103/PhysRevLett.108.109903} {\bibfield  {journal} {\bibinfo  {journal}
  {Phys. Rev. Lett.}\ }\textbf {\bibinfo {volume} {108}},\ \bibinfo {pages}
  {109903} (\bibinfo {year} {2012})}\BibitemShut {NoStop}%
\bibitem [{\citenamefont {Lepers}, \citenamefont {Guillon},\ and\ \citenamefont
  {Honvault}(2019)}]{LepersMaxence2019Qmso}%
  \BibitemOpen
  \bibfield  {author} {\bibinfo {author} {\bibfnamefont {M.}~\bibnamefont
  {Lepers}}, \bibinfo {author} {\bibfnamefont {G.}~\bibnamefont {Guillon}}, \
  and\ \bibinfo {author} {\bibfnamefont {P.}~\bibnamefont {Honvault}},\
  }\href@noop {} {\bibfield  {journal} {\bibinfo  {journal} {Monthly Notices of
  the Royal Astronomical Society}\ }\textbf {\bibinfo {volume} {488}},\
  \bibinfo {pages} {4732--4739} (\bibinfo {year} {2019})}\BibitemShut {NoStop}%
\bibitem [{\citenamefont {Lepers}, \citenamefont {Guillon},\ and\ \citenamefont
  {Honvault}(2020)}]{10.1093/mnras/staa3146}%
  \BibitemOpen
  \bibfield  {author} {\bibinfo {author} {\bibfnamefont {M.}~\bibnamefont
  {Lepers}}, \bibinfo {author} {\bibfnamefont {G.}~\bibnamefont {Guillon}}, \
  and\ \bibinfo {author} {\bibfnamefont {P.}~\bibnamefont {Honvault}},\ }\href
  {\doibase 10.1093/mnras/staa3146} {\bibfield  {journal} {\bibinfo  {journal}
  {Monthly Notices of the Royal Astronomical Society}\ }\textbf {\bibinfo
  {volume} {500}},\ \bibinfo {pages} {430--431} (\bibinfo {year}
  {2020})}\BibitemShut {NoStop}%
\bibitem [{\citenamefont {Desrousseaux}\ \emph {et~al.}(2021)\citenamefont
  {Desrousseaux}, \citenamefont {Konings}, \citenamefont {Loreau},\ and\
  \citenamefont {Lique}}]{Desrousseaux}%
  \BibitemOpen
  \bibfield  {author} {\bibinfo {author} {\bibfnamefont {B.}~\bibnamefont
  {Desrousseaux}}, \bibinfo {author} {\bibfnamefont {M.}~\bibnamefont
  {Konings}}, \bibinfo {author} {\bibfnamefont {J.}~\bibnamefont {Loreau}}, \
  and\ \bibinfo {author} {\bibfnamefont {F.}~\bibnamefont {Lique}},\ }\href
  {\doibase 10.1039/D1CP02564C} {\bibfield  {journal} {\bibinfo  {journal}
  {Phys. Chem. Chem. Phys.}\ } (\bibinfo {year} {2021}),\
  10.1039/D1CP02564C}\BibitemShut {NoStop}%
\bibitem [{\citenamefont {Zanchet}\ \emph {et~al.}(2019)\citenamefont
  {Zanchet}, \citenamefont {Lique}, \citenamefont {Roncero}, \citenamefont
  {Goicoechea},\ and\ \citenamefont {Bulut}}]{Zanchet_2019}%
  \BibitemOpen
  \bibfield  {author} {\bibinfo {author} {\bibfnamefont {A.}~\bibnamefont
  {Zanchet}}, \bibinfo {author} {\bibfnamefont {F.}~\bibnamefont {Lique}},
  \bibinfo {author} {\bibfnamefont {O.}~\bibnamefont {Roncero}}, \bibinfo
  {author} {\bibfnamefont {J.}~\bibnamefont {Goicoechea}}, \ and\ \bibinfo
  {author} {\bibfnamefont {N.}~\bibnamefont {Bulut}},\ }\href@noop {}
  {\bibfield  {journal} {\bibinfo  {journal} {Astronomy \& Astrophysics}\
  }\textbf {\bibinfo {volume} {626}} (\bibinfo {year} {2019})}\BibitemShut
  {NoStop}%
\bibitem [{\citenamefont {{Lique, Fran\c{c}ois}}\ \emph
  {et~al.}(2020)\citenamefont {{Lique, Fran\c{c}ois}}, \citenamefont {{Zanchet,
  Alexandre}}, \citenamefont {{Bulut, Niyazi}}, \citenamefont {{Goicoechea,
  Javier R.}},\ and\ \citenamefont {{Roncero, Octavio}}}]{refId0}%
  \BibitemOpen
  \bibfield  {author} {\bibinfo {author} {\bibnamefont {{Lique,
  Fran\c{c}ois}}}, \bibinfo {author} {\bibnamefont {{Zanchet, Alexandre}}},
  \bibinfo {author} {\bibnamefont {{Bulut, Niyazi}}}, \bibinfo {author}
  {\bibnamefont {{Goicoechea, Javier R.}}}, \ and\ \bibinfo {author}
  {\bibnamefont {{Roncero, Octavio}}},\ }\href {\doibase
  10.1051/0004-6361/202038041} {\bibfield  {journal} {\bibinfo  {journal}
  {A\&A}\ }\textbf {\bibinfo {volume} {638}},\ \bibinfo {pages} {A72} (\bibinfo
  {year} {2020})}\BibitemShut {NoStop}%
\bibitem [{\citenamefont {Faure}\ and\ \citenamefont
  {Lique}(2012)}]{Faure&Lique_IOS}%
  \BibitemOpen
  \bibfield  {author} {\bibinfo {author} {\bibfnamefont {A.}~\bibnamefont
  {Faure}}\ and\ \bibinfo {author} {\bibfnamefont {F.}~\bibnamefont {Lique}},\
  }\href {\doibase https://doi.org/10.1111/j.1365-2966.2012.21601.x} {\bibfield
   {journal} {\bibinfo  {journal} {Monthly Notices of the Royal Astronomical
  Society}\ }\textbf {\bibinfo {volume} {425}},\ \bibinfo {pages} {740--748}
  (\bibinfo {year} {2012})}\BibitemShut {NoStop}%
\bibitem [{\citenamefont {Zanchet}\ \emph {et~al.}(2013)\citenamefont
  {Zanchet}, \citenamefont {Ag{\'{u}}ndez}, \citenamefont {Herrero},
  \citenamefont {Aguado},\ and\ \citenamefont {Roncero}}]{Zanchet_2013}%
  \BibitemOpen
  \bibfield  {author} {\bibinfo {author} {\bibfnamefont {A.}~\bibnamefont
  {Zanchet}}, \bibinfo {author} {\bibfnamefont {M.}~\bibnamefont
  {Ag{\'{u}}ndez}}, \bibinfo {author} {\bibfnamefont {V.~J.}\ \bibnamefont
  {Herrero}}, \bibinfo {author} {\bibfnamefont {A.}~\bibnamefont {Aguado}}, \
  and\ \bibinfo {author} {\bibfnamefont {O.}~\bibnamefont {Roncero}},\ }\href
  {\doibase 10.1088/0004-6256/146/5/125} {\bibfield  {journal} {\bibinfo
  {journal} {The Astronomical Journal}\ }\textbf {\bibinfo {volume} {146}},\
  \bibinfo {pages} {125} (\bibinfo {year} {2013})}\BibitemShut {NoStop}%
\bibitem [{\citenamefont {Werfelli}\ \emph {et~al.}(2015)\citenamefont
  {Werfelli}, \citenamefont {Halvick}, \citenamefont {Honvault}, \citenamefont
  {Kerkeni},\ and\ \citenamefont {Stoecklin}}]{WerfelliGhofran2015Ltrc}%
  \BibitemOpen
  \bibfield  {author} {\bibinfo {author} {\bibfnamefont {G.}~\bibnamefont
  {Werfelli}}, \bibinfo {author} {\bibfnamefont {P.}~\bibnamefont {Halvick}},
  \bibinfo {author} {\bibfnamefont {P.}~\bibnamefont {Honvault}}, \bibinfo
  {author} {\bibfnamefont {B.}~\bibnamefont {Kerkeni}}, \ and\ \bibinfo
  {author} {\bibfnamefont {T.}~\bibnamefont {Stoecklin}},\ }\href@noop {}
  {\bibfield  {journal} {\bibinfo  {journal} {The Journal of Chemical Physics}\
  }\textbf {\bibinfo {volume} {143}} (\bibinfo {year} {2015})}\BibitemShut
  {NoStop}%
\bibitem [{\citenamefont {Plasil}\ \emph {et~al.}(2011)\citenamefont {Plasil},
  \citenamefont {Mehner}, \citenamefont {Dohnal}, \citenamefont {Kotrik},
  \citenamefont {Glosik},\ and\ \citenamefont {Gerlich}}]{Plasil_2011}%
  \BibitemOpen
  \bibfield  {author} {\bibinfo {author} {\bibfnamefont {R.}~\bibnamefont
  {Plasil}}, \bibinfo {author} {\bibfnamefont {T.}~\bibnamefont {Mehner}},
  \bibinfo {author} {\bibfnamefont {P.}~\bibnamefont {Dohnal}}, \bibinfo
  {author} {\bibfnamefont {T.}~\bibnamefont {Kotrik}}, \bibinfo {author}
  {\bibfnamefont {J.}~\bibnamefont {Glosik}}, \ and\ \bibinfo {author}
  {\bibfnamefont {D.}~\bibnamefont {Gerlich}},\ }\href {\doibase
  10.1088/0004-637x/737/2/60} {\bibfield  {journal} {\bibinfo  {journal} {The
  Astrophysical Journal}\ }\textbf {\bibinfo {volume} {737}},\ \bibinfo {pages}
  {60} (\bibinfo {year} {2011})}\BibitemShut {NoStop}%
\bibitem [{\citenamefont {Warmbier}\ and\ \citenamefont
  {Schneider}(2011)}]{Warmbier}%
  \BibitemOpen
  \bibfield  {author} {\bibinfo {author} {\bibfnamefont {R.}~\bibnamefont
  {Warmbier}}\ and\ \bibinfo {author} {\bibfnamefont {R.}~\bibnamefont
  {Schneider}},\ }\href {\doibase 10.1039/C1CP20212J} {\bibfield  {journal}
  {\bibinfo  {journal} {Phys. Chem. Chem. Phys.}\ }\textbf {\bibinfo {volume}
  {13}},\ \bibinfo {pages} {10285--10294} (\bibinfo {year} {2011})}\BibitemShut
  {NoStop}%
\bibitem [{\citenamefont {Grozdanov}\ and\ \citenamefont
  {McCarroll}(2013)}]{GROZDANOV201323}%
  \BibitemOpen
  \bibfield  {author} {\bibinfo {author} {\bibfnamefont {T.}~\bibnamefont
  {Grozdanov}}\ and\ \bibinfo {author} {\bibfnamefont {R.}~\bibnamefont
  {McCarroll}},\ }\href {\doibase https://doi.org/10.1016/j.cplett.2013.04.076}
  {\bibfield  {journal} {\bibinfo  {journal} {Chemical Physics Letters}\
  }\textbf {\bibinfo {volume} {575}},\ \bibinfo {pages} {23--26} (\bibinfo
  {year} {2013})}\BibitemShut {NoStop}%
\bibitem [{\citenamefont {Li}, \citenamefont {Zhang},\ and\ \citenamefont
  {Han}(2015)}]{Li}%
  \BibitemOpen
  \bibfield  {author} {\bibinfo {author} {\bibfnamefont {Y.~Q.}\ \bibnamefont
  {Li}}, \bibinfo {author} {\bibfnamefont {P.~Y.}\ \bibnamefont {Zhang}}, \
  and\ \bibinfo {author} {\bibfnamefont {K.~L.}\ \bibnamefont {Han}},\ }\href
  {\doibase 10.1063/1.4916035} {\bibfield  {journal} {\bibinfo  {journal} {The
  Journal of Chemical Physics}\ }\textbf {\bibinfo {volume} {142}},\ \bibinfo
  {pages} {124302} (\bibinfo {year} {2015})}\BibitemShut {NoStop}%
\bibitem [{\citenamefont {Bovino}, \citenamefont {Grassi},\ and\ \citenamefont
  {Gianturco}(2015)}]{Bovino}%
  \BibitemOpen
  \bibfield  {author} {\bibinfo {author} {\bibfnamefont {S.}~\bibnamefont
  {Bovino}}, \bibinfo {author} {\bibfnamefont {T.}~\bibnamefont {Grassi}}, \
  and\ \bibinfo {author} {\bibfnamefont {F.~A.}\ \bibnamefont {Gianturco}},\
  }\href {\doibase 10.1021/acs.jpca.5b02785} {\bibfield  {journal} {\bibinfo
  {journal} {The Journal of Physical Chemistry A}\ }\textbf {\bibinfo {volume}
  {119}},\ \bibinfo {pages} {11973--11982} (\bibinfo {year}
  {2015})}\BibitemShut {NoStop}%
\bibitem [{\citenamefont {Skouteris}, \citenamefont {Castillo},\ and\
  \citenamefont {Manolopoulos}(2000)}]{Skouteris2000}%
  \BibitemOpen
  \bibfield  {author} {\bibinfo {author} {\bibfnamefont {D.}~\bibnamefont
  {Skouteris}}, \bibinfo {author} {\bibfnamefont {J.}~\bibnamefont {Castillo}},
  \ and\ \bibinfo {author} {\bibfnamefont {D.~E.}\ \bibnamefont
  {Manolopoulos}},\ }\href@noop {} {\bibfield  {journal} {\bibinfo  {journal}
  {Comput. Phys. Comm.}\ }\textbf {\bibinfo {volume} {133}},\ \bibinfo {pages}
  {128} (\bibinfo {year} {2000})}\BibitemShut {NoStop}%
\bibitem [{\citenamefont {Nuñez-Reyes}\ \emph {et~al.}(2018)\citenamefont
  {Nuñez-Reyes}, \citenamefont {Hickson}, \citenamefont {Larrégaray},
  \citenamefont {Bonnet}, \citenamefont {González-Lezana},\ and\ \citenamefont
  {Suleimanov}}]{Reyes}%
  \BibitemOpen
  \bibfield  {author} {\bibinfo {author} {\bibfnamefont {D.}~\bibnamefont
  {Nuñez-Reyes}}, \bibinfo {author} {\bibfnamefont {K.~M.}\ \bibnamefont
  {Hickson}}, \bibinfo {author} {\bibfnamefont {P.}~\bibnamefont
  {Larrégaray}}, \bibinfo {author} {\bibfnamefont {L.}~\bibnamefont {Bonnet}},
  \bibinfo {author} {\bibfnamefont {T.}~\bibnamefont {González-Lezana}}, \
  and\ \bibinfo {author} {\bibfnamefont {Y.~V.}\ \bibnamefont {Suleimanov}},\
  }\href {\doibase 10.1039/C7CP07843A} {\bibfield  {journal} {\bibinfo
  {journal} {Phys. Chem. Chem. Phys.}\ }\textbf {\bibinfo {volume} {20}},\
  \bibinfo {pages} {4404--4414} (\bibinfo {year} {2018})}\BibitemShut {NoStop}%
\bibitem [{\citenamefont {Nuñez-Reyes}\ \emph {et~al.}(2020)\citenamefont
  {Nuñez-Reyes}, \citenamefont {Bray}, \citenamefont {Hickson}, \citenamefont
  {Larrégaray}, \citenamefont {Bonnet},\ and\ \citenamefont
  {González-Lezana}}]{Reyes2}%
  \BibitemOpen
  \bibfield  {author} {\bibinfo {author} {\bibfnamefont {D.}~\bibnamefont
  {Nuñez-Reyes}}, \bibinfo {author} {\bibfnamefont {C.}~\bibnamefont {Bray}},
  \bibinfo {author} {\bibfnamefont {K.~M.}\ \bibnamefont {Hickson}}, \bibinfo
  {author} {\bibfnamefont {P.}~\bibnamefont {Larrégaray}}, \bibinfo {author}
  {\bibfnamefont {L.}~\bibnamefont {Bonnet}}, \ and\ \bibinfo {author}
  {\bibfnamefont {T.}~\bibnamefont {González-Lezana}},\ }\href {\doibase
  10.1039/D0CP03971C} {\bibfield  {journal} {\bibinfo  {journal} {Phys. Chem.
  Chem. Phys.}\ }\textbf {\bibinfo {volume} {22}},\ \bibinfo {pages}
  {23609--23617} (\bibinfo {year} {2020})}\BibitemShut {NoStop}%
\bibitem [{\citenamefont {González-Lezana}, \citenamefont {Larrégaray},\ and\
  \citenamefont {Bonnet}(2021)}]{GONZALEZLEZANA2021138228}%
  \BibitemOpen
  \bibfield  {author} {\bibinfo {author} {\bibfnamefont {T.}~\bibnamefont
  {González-Lezana}}, \bibinfo {author} {\bibfnamefont {P.}~\bibnamefont
  {Larrégaray}}, \ and\ \bibinfo {author} {\bibfnamefont {L.}~\bibnamefont
  {Bonnet}},\ }\href {\doibase https://doi.org/10.1016/j.cplett.2020.138228}
  {\bibfield  {journal} {\bibinfo  {journal} {Chemical Physics Letters}\
  }\textbf {\bibinfo {volume} {763}},\ \bibinfo {pages} {138228} (\bibinfo
  {year} {2021})}\BibitemShut {NoStop}%
\bibitem [{\citenamefont {González-Lezana}\ \emph {et~al.}(2018)\citenamefont
  {González-Lezana}, \citenamefont {Larrégaray}, \citenamefont {Bonnet},
  \citenamefont {Wu},\ and\ \citenamefont {Bian}}]{Yanan}%
  \BibitemOpen
  \bibfield  {author} {\bibinfo {author} {\bibfnamefont {T.}~\bibnamefont
  {González-Lezana}}, \bibinfo {author} {\bibfnamefont {P.}~\bibnamefont
  {Larrégaray}}, \bibinfo {author} {\bibfnamefont {L.}~\bibnamefont {Bonnet}},
  \bibinfo {author} {\bibfnamefont {Y.}~\bibnamefont {Wu}}, \ and\ \bibinfo
  {author} {\bibfnamefont {W.}~\bibnamefont {Bian}},\ }\href {\doibase
  10.1063/1.5026454} {\bibfield  {journal} {\bibinfo  {journal} {The Journal of
  Chemical Physics}\ }\textbf {\bibinfo {volume} {148}},\ \bibinfo {pages}
  {234305} (\bibinfo {year} {2018})}\BibitemShut {NoStop}%
\bibitem [{\citenamefont {Nuñez-Reyes}\ \emph {et~al.}(2019)\citenamefont
  {Nuñez-Reyes}, \citenamefont {Hickson}, \citenamefont {Larrégaray},
  \citenamefont {Bonnet}, \citenamefont {González-Lezana}, \citenamefont
  {Bhowmick},\ and\ \citenamefont {Suleimanov}}]{Reyes3}%
  \BibitemOpen
  \bibfield  {author} {\bibinfo {author} {\bibfnamefont {D.}~\bibnamefont
  {Nuñez-Reyes}}, \bibinfo {author} {\bibfnamefont {K.~M.}\ \bibnamefont
  {Hickson}}, \bibinfo {author} {\bibfnamefont {P.}~\bibnamefont
  {Larrégaray}}, \bibinfo {author} {\bibfnamefont {L.}~\bibnamefont {Bonnet}},
  \bibinfo {author} {\bibfnamefont {T.}~\bibnamefont {González-Lezana}},
  \bibinfo {author} {\bibfnamefont {S.}~\bibnamefont {Bhowmick}}, \ and\
  \bibinfo {author} {\bibfnamefont {Y.~V.}\ \bibnamefont {Suleimanov}},\ }\href
  {\doibase 10.1021/acs.jpca.9b06133} {\bibfield  {journal} {\bibinfo
  {journal} {The Journal of Physical Chemistry A}\ }\textbf {\bibinfo {volume}
  {123}},\ \bibinfo {pages} {8089--8098} (\bibinfo {year} {2019})}\BibitemShut
  {NoStop}%
\end{thebibliography}%
	\end{document}